\documentclass[preprint, amsmath, amssymb, aps]{revtex4-2}
\pdfoutput=1
\usepackage{graphicx,graphics}
\usepackage[export]{adjustbox}
\usepackage{verbatim}
\usepackage[font=small]{caption}
\usepackage{braket}
\usepackage{array}
\usepackage{float}
\usepackage{breqn}
\usepackage{subfig}
\usepackage{color}
\usepackage[colorlinks=true, linkcolor=blue, citecolor=blue, urlcolor=blue]{hyperref}

\usepackage{wrapfig}
\usepackage[utf8]{inputenc}
\usepackage{booktabs} 
\usepackage{multirow} 

\usepackage{rotating}
\usepackage[normalem]{ulem}
\usepackage{enumerate}
\makeatletter
\let\cat@comma@active\@empty
\makeatother
\begin{document}
	
    \title{\textcolor{blue}{Emergence of Berry curvature dipole and nonlinear Hall effect in a novel hexagonal borophosphene allotrope}
}
    
      \author{Ranjan Kumar Tung} 
      \thanks{}
      \affiliation{Department of Physics, GITAM School of Science, GITAM (Deemed to be University), Bengaluru 561203, Karnataka, India}
      \author{Dimple} 
      \thanks{dimple@gitam.edu}
      \affiliation{Theoretical and Computational Modeling of Materials Lab (TCM2),Department of Physics, GITAM School of Science, GITAM (Deemed to be University), Bengaluru 561203, Karnataka, India}
      \author{Venkatesan Subramanian}
      \thanks{}
      \affiliation{Department of Chemistry, Indian Institute of Technology (IIT) Madras,Chennai 600036, India}
      \affiliation{Centre for Advanced Computational and Theoretical Science (ACTS), SRM Institute of Science and Technology, Kattankulathur - 603203, Tamil Naidu, India }
       \affiliation{Department of Chemistry, SRM Institute of Science and Technology, Kattankulathur - 603203, Tamil Naidu, India }
      \author{Vijay Singh} 
      \thanks{vsingh2@gitam.edu}
      \affiliation{Department of Physics, GITAM School of Science, GITAM (Deemed to be University), Bengaluru 561203, Karnataka, India}

\begin{abstract}
We investigate the electronic structure, local orbital coupling, and strain-tunable nonlinear Hall response of two arsenic-substituted borophosphene-derived monolayers, As@MBP and As@AMBP, using first-principles calculations and tight-binding analysis. In the absence of spin--orbit coupling, both systems exhibit a finite Berry-curvature dipole (BCD) and a substantial nonlinear Hall conductivity, demonstrating that the nonlinear response is intrinsically driven by orbital physics and inversion-symmetry breaking. $As$ substitution opens sizable band gaps of $120~\mathrm{meV}$ in As@MBP and $90~\mathrm{meV}$ in As@AMBP, well exceeding room-temperature thermal fluctuations ($k_{\mathrm{B}}T \approx 25.9~\mathrm{meV}$). Tight-binding analysis reveals that $As$ incorporation breaks the equivalence of $B-P$ hoppings in As@MBP and induces sign reversals in key nearest-neighbor hoppings in As@AMBP, reflecting a pronounced electronic reconstruction beyond simple geometric effects. Under mechanical strain, the redistribution of Berry curvature yields a maximum BCD of $22~\mathrm{\AA}$ in As@MBP, approximately $10$ times larger than that of unstrained structure, and induces a reversal of the BCD sign. These results establish $As$-substituted borophosphene derivatives as promising light-element platforms for mechanically switchable nonlinear transport and quantum-material applications without relying on strong spin-orbit coupling.
\end{abstract}


      \maketitle

 \section{Introduction}
 
The geometry and topology of electronic wave functions provide a powerful framework for understanding a wide range of emergent quantum phenomena~\cite{Berry,Zoo:Topo,Haldane-Nobel}. These concepts underlie the quantum Hall effect~\cite{QHE}, anomalous transport~\cite{AHE}, and topological phases of matter~\cite{z2,kane-Mele,Fu-Kane-mele,Topo-Ins-Super}. In crystalline solids, this geometry is captured by the Berry curvature $\boldsymbol{\Omega}(\mathbf{k})$, a gauge-invariant quantity defined in momentum space~\cite{Coll:hasan-Kane,AHE-Nagaosa,TopBanTh}. Physically, $\boldsymbol{\Omega}(\mathbf{k})$ is the geometric phase acquired by a Bloch state per unit area when it is adiabatically transported around a closed loop in the Brillouin zone (BZ). It acts as an effective magnetic field in momentum space and produces an anomalous velocity in the semiclassical equations of motion, $\dot{\mathbf{r}} = \hbar^{-1}\nabla_{\mathbf{k}}\epsilon_n(\mathbf{k}) - \dot{\mathbf{k}}\times\boldsymbol{\Omega}(\mathbf{k})$, which governs a variety of unconventional transport responses in quantum materials~\cite{berry-mag-fld,AHE-Nagaosa}.



A prominent manifestation of Berry-curvature-driven transport is the anomalous Hall effect (AHE) \cite{Anomalous}, which occurs in systems with broken time-reversal symmetry ($\mathcal{T}$). The resulting finite Berry curvature produces a transverse Hall current that is linear in the applied electric field. In such systems, the interplay between an external electric field and a finite Berry curvature generates a transverse Hall current even in the absence of an external magnetic field \cite{AHE}. This Hall response is linear in the applied electric field and serves as a hallmark of the intrinsic geometric properties of the electronic band structure.

Traditionally, Hall responses have been understood within the framework of linear response theory, where a transverse voltage is generated by external electric or magnetic fields \cite{AHE-Nagaosa,AHE,Anomalous}. The observation of conventional and anomalous Hall effects generally requires broken $\mathcal{T}$, induced either by an external magnetic field or intrinsic magnetic ordering. In contrast, the second-order nonlinear Hall effect (NLHE) \cite{Inti} can emerge in $\mathcal{T}$-symmetric nonmagnetic systems, provided spatial inversion symmetry ($\mathcal{P}$) is broken, typically mediated by a non-zero Berry curvature dipole (BCD). 
Unlike the linear Hall effect, which is governed by the Berry curvature itself, the intrinsic nonlinear Hall response originates from the BCD  , defined as the first moment of the Berry curvature distribution across occupied electronic states \cite{Du2021NonlinearHE,Awadesh-review}. This quantifies the asymmetry of the Berry-curvature distribution in the Brillouin zone and gives rise to a transverse current that is quadratic in the applied electric field\cite{Inti}. This discovery has established a new paradigm for Hall transport beyond the conventional linear-response regime\cite{Binghai,Joseph-awadesh}. Additionally, higher-order moments, such as the Berry-curvature quadrupole, have been predicted to govern higher-order nonlinear transport phenomena \cite{HONLHE,3rd-NLHE}. Owing to their potential applications in rectification, frequency doubling, terahertz detection, and symmetry-controlled electronic devices, the search for materials hosting large and robust BCDs remains an active frontier in condensed matter physics \cite{Awadesh-review,Joseph-awadesh}. 



It is important to note that the nonlinear Hall effect can also originate from the full quantum geometry of Bloch electrons in $\mathbf{k}$-space, encoded in the quantum geometric tensor (QGT). Its imaginary part gives the Berry curvature, $\mathrm{Im}(\mathrm{QGT})$, which tracks how the phase of a state winds along a closed loop. Its real part gives the quantum metric, $\mathrm{Re}(\mathrm{QGT})$, which measures how rapidly a state changes (the wave-function overlap) when moving from $\mathbf{k}$ to $\mathbf{k}+d\mathbf{k}$. In $\mathcal{PT}$-symmetric systems the Berry curvature vanishes identically, so the BCD is zero, but the quantum metric survives. The quantum metric dipole (QMD), the momentum-space inhomogeneity of the quantum metric, then provides a geometric correction to the velocity and enables an intrinsic nonlinear Hall response, as demonstrated in the layered antiferromagnet $\mathrm{MnBi_2Te_4}$~\cite{Metric,quan-metric,QGT,Amit-QMD,QMD,Amit-NLVHE,MB-QMD,Altmag-nlhe,mnbitblkBp,Mag-quan-geo,MnBiTe4-Binghai}.


Theoretical and experimental studies have identified several classes of materials exhibiting sizable BCDs, including transition-metal dichalcogenides, such as $WTe_2$, $MoTe_2$ and $WSe_2$, where inversion symmetry can be manipulated through strain, gating, or layer engineering \cite{WTe2-MoTe2,GiantWSe2,Awadesh-review, jena2019valley}. Large nonlinear Hall responses have also been reported in graphene-based systems \cite{FSWarping-graphene,fermi-surf}, where Fermi-surface anisotropy and hexagonal warping generate substantial BCDs, further enhanced in twisted bilayer graphene\cite{TBG}. Beyond low-dimensional platforms, three-dimensional topological semimetals such as bulk $\mathrm{WTe_2}$~\cite{Ma2019_WTe2_NLHE}, $\mathrm{Cd_3As_2}$~\cite{nonlinear_Cd3As2}, $\mathrm{TaIrTe_4}$~\cite{TaIrTe4}, $\mathrm{TaAs}$~\cite{TaAs_tunable_bcd}, and non-centrosymmetric Weyl systems like $\mathrm{Ce_3Bi_4Pd_3}$~\cite{CeBiPd} have emerged as viable hosts for nonlinear Hall transport~\cite{3D-NLHmaterial, Ma_2018, Kang2019NonlinearAH, nlhe-ortix}, alongside recent experimental demonstrations in van der Waals bilayers~\cite{hetroNLHE, hetro-superlattice} and ferroelectric heterostructures~\cite{ferro-NLH}. \\
Along these lines, two-dimensional systems such as borophene and black phosphorus, with their distinctive electronic structures and symmetry properties, provide a compelling platform for investigating symmetry-driven transport phenomena.~\cite{Molle2017, Liu2014} However, their practical deployment as low-dimensional nonlinear elements remains severely constrained by environmental degradation~\cite{Favron2015, Castellanos2015} and complex structural polymorphism~\cite{Mannix2015}. Binary Group III--V monolayers overcome these intrinsic vulnerabilities~\cite{Yagmurcukardes2020}, positioning monolayer boron phosphide (2D BP / borophosphene) - a honeycomb or orthorhombic lattice of alternating B and P atoms - as a exceptionally versatile platform~\cite{Zheng2016}. Crucially, strong covalent B--P hybridization passivates phosphorus lone-pair electrons to ensure superior ambient stability~\cite{Zhang2019}, while its electronic structure spans from an indirect-gap semiconductor ($\sim 0.90~\mathrm{eV}$)~\cite{Li2021} to a robust 2D Dirac material hosting anisotropic, distorted Dirac cones with ultrahigh Fermi velocities ($\sim 10^5~\mathrm{m/s}$)~\cite{Zhang2019}. Beyond its established utility in high-performance energy storage (exhibiting capacities up to $1282.7~\mathrm{mA\,h\,g^{-1}}$ and diffusion barriers as low as $0.12-0.14~\mathrm{eV}$)~\cite{Wang2020, Esackraj2023} and thermoelectrics ($ZT = 0.255$)~\cite{Li2021}, the combination of structural stability, highly anisotropic band dispersion, and lightweight elemental constituent makes 2D boron phosphide a fertile yet unexplored ground for mechanical symmetry tuning and geometric nonlinear transport.

In this study, we examine the electronic structures of two centrosymmetric materials: modified borophosphene (MBP) and acetylene-mediated borophosphene (AMBP). To facilitate nonlinear transport, we propose two lower-symmetry derivatives, As@MBP and As@AMBP, engineered by strategically substituting arsenic atoms to break spatial inversion symmetry. Density functional theory calculations reveal that this symmetry reduction lifts the Dirac point degeneracy, opening a trivial energy band gap for As@MBP and As@AMBP. Using a tight-binding Hamiltonian formulation, we evaluate $p_z-p_z$ nearest-neighbor hopping integrals across adjacent sites to track variations in their electronic dispersion. By calculating the first moment of the Berry curvature - defined as the BCD  - we quantify the intrinsic non-magnetic, time-reversal-invariant nonlinear Hall response in these derivatives. Furthermore, we demonstrate that mechanical strain serves as a powerful tuning knob, enabling precise control over the bandgap, driving up to a tenfold amplification in BCD magnitude, and enabling sign-reversal modulation.

The roadmap of this paper is structured as follows: first, we outline the mathematical formalism governing the origin of the BCD-driven nonlinear Hall effect in $\mathcal{T}$ invariant systems. Next, we present the crystal structures of pristine MBP/AMBP and their $\mathcal{P}$-broken As-substituted derivatives. We then describe the first-principles and tight-binding computational methodologies used in this work. Finally, we discuss the electronic dispersion, strain-engineered Berry curvature redistribution, and BCD transport results, before concluding with the topological and practical implications of these SOC-free borophosphene derivatives.

\section{From Linear to Nonlinear Hall Effect}

\subsubsection{The Origin of the Anomalous Hall Effect (AHE)}
The anomalous Hall effect is a fundamental transport phenomenon rooted in the intrinsic Berry curvature $\mathbf{\Omega}_n(\mathbf{k})$ of the occupied electronic bands. In systems where $\mathcal{T}$ is broken, such as ferromagnets or magnetic topological insulators, a transverse Hall voltage is generated even in the absence of an external magnetic field. The semiclassical equations of motion for a Bloch electron under in-plane time-dependent electric field $\mathbf{E}(t)$ include an anomalous velocity term \cite{AHE-Nagaosa,Anomalous,Inti}:
\begin{equation}
\dot{\mathbf{r}} = \mathbf{v}_n(\mathbf{k}) - \dot{\mathbf{k}} \times \Omega_n^c(\mathbf{k})\hat{z}
\end{equation}
\begin{equation}
\hbar \dot{\mathbf{k}} = -e\mathbf{E}(t)
\end{equation}
where $\mathbf{v}_n(\mathbf{k}) = \frac{1}{\hbar}\boldsymbol{\nabla}_{\mathbf{k}} \epsilon_n(\mathbf{k})$ is the unperturbed group velocity of band $n$, and $e$ is the elementary charge. The anomalous velocity component perpendicular to the electric field gives rise to the linear anomalous Hall current density \cite{Anomalous}:
\begin{equation}
\mathbf{j}_{\text{AH}} = -\frac{e^2}{\hbar} \sum_{n} \int \frac{d^2k}{(2\pi)^2} f_n(\mathbf{k}) [\mathbf{E}\times \Omega_n^c(\mathbf{k})\hat{z}]
\end{equation}
The resulting linear Hall conductivity tensor component $\sigma_{ab}^{\text{AHE}}$ is determined by integrating the Berry curvature $\Omega_{n}^{c}(\mathbf{k})$ over n-occupied states \cite{AHE}:
\begin{equation}
\sigma_{ab}^{\text{AH}} = -\frac{e^2}{\hbar} \epsilon_{ab} \sum_{n} \int \frac{d^2k}{(2\pi)^2} f_n(\mathbf{k})\Omega_{n}^{c}(\mathbf{k})
\end{equation}
where $\epsilon_{ab}$ is the 2D Levi-Civita antisymmetric tensor, and $f_n(\mathbf{k})$ is the Fermi-Dirac distribution function for band $n$. The out-of-plane component of the Berry curvature $\Omega_{n}^{c}(\mathbf{k})$ can be evaluated using the Kubo formula \cite{kubo,Vanderbilt}:
\begin{equation}\label{BC}
\Omega_{n}^{c}(\mathbf{k}) = -2\hbar^2 Im \sum_{m \neq n} \frac{\langle n,\mathbf{k} | \hat{v}_a | m,\mathbf{k} \rangle \langle m,\mathbf{k} | \hat{v}_b | n,\mathbf{k} \rangle}{\left[ E_m(\mathbf{k}) - E_n(\mathbf{k}) \right]^2}
\end{equation}
where $\hat{v}_i$ represent the velocity operators along orthogonal spatial directions.

\subsubsection{Transition to the Nonlinear Hall Effect (NLHE)}

To observe a pure nonlinear transport response, one must suppress the linear anomalous Hall effect (AHE). This is achieved in systems that preserve $\mathcal{T}$ symmetry, which forces the global Berry curvature to be an odd function in momentum space, $\mathbf{\Omega}^c_n(-\mathbf{k}) = -\mathbf{\Omega}^c_n(\mathbf{k})$, causing the total linear Hall conductivity to vanish. However, if  $\mathcal{P}$ symmetry is simultaneously broken, local non-zero Berry curvature $\mathbf{\Omega}^c_n(\mathbf{k})$ is permitted. 

Under an applied electric field, the out-of-equilibrium electronic distribution $f_n(\mathbf{k}, t)$ for band $n$ is governed by the Boltzmann transport equation (BTE) within the relaxation-time approximation \cite{kubo}:
\begin{equation}
\frac{\partial f_n(\mathbf{k}, t)}{\partial t} + \dot{\mathbf{k}} \cdot \boldsymbol{\nabla}_{\mathbf{k}} f_n(\mathbf{k}, t) = -\frac{f_n(\mathbf{k}, t) - f_n^0(\mathbf{k})}{\tau}
\end{equation}
where $\tau$ is the intra-band relaxation time and $f_n^0(\mathbf{k}) = \left[ 1 + \exp\left( \frac{\epsilon_n(\mathbf{k}) - \mu}{k_B T} \right) \right]^{-1}$ is the equilibrium Fermi-Dirac distribution function. For a monochromatic AC driving field $\mathbf{E}(t) = \mathbf{E}_{\omega} e^{-i\omega t} + \text{c.c.}$, the linear first-order perturbation to the distribution function $f_n^{(1)}(\mathbf{k}, \omega)$ is given by:
\begin{equation}
\begin{split}
f_n^{(1)}(\mathbf{k}, \omega) = \frac{e\tau}{\hbar(1 + i\omega\tau)} \left( \mathbf{E}_{\omega} \cdot \boldsymbol{\nabla}_{\mathbf{k}} f_n^0(\mathbf{k}) \right) \\= \frac{e\tau}{\hbar(1 + i\omega\tau)} \left( \mathbf{E}_{\omega} \cdot \mathbf{v}_n(\mathbf{k}) \right) \frac{\partial f_n^0}{\partial \epsilon_n}
\end{split}
\end{equation}

The second-order transverse current density $\mathbf{j}^{(2)}$ arises from the coupling of the anomalous velocity term with the non-equilibrium distribution function $f_n^{(1)}(\mathbf{k}, t)$. Under $\mathcal{T}$-symmetry, the contribution from the symmetric group velocity vanishes upon integration over the Brillouin zone, leaving the anomalous velocity as the sole driver of the intrinsic nonlinear Hall current \cite{Inti}:
\begin{equation}
j_a^{(2)} = -e \sum_{n} \int \frac{d^2k}{(2\pi)^2} f_n^{(1)}(\mathbf{k}, t) \left[- \frac{e}{\hbar} \epsilon_{ab} E_b(t) \Omega_n^c(\mathbf{k}) \right]
\end{equation}

This non-linear interaction generates both a rectified direct current (DC) response and a frequency-doubled second-harmonic ($2\omega$) signal. The dc part of nonlinear Hall conductivity tensor $\chi_{abd}$, defined via $j_a^{(2)} = \chi_{abd} E_b E_d^*$, is expressed as\cite{Inti}:
\begin{equation}
\chi_{abd} = -\epsilon_{ab} \frac{e^3 \tau}{2\hbar^2 (1 +\omega^2\tau^2)} D_{d}
\end{equation}
where $D_{d}$ is the BCD defined over all occupied states and d $\in$  {a,b}.

\subsubsection{Berry Curvature Dipole}

The BCD provides an important measure of the
momentum-space asymmetry of the Berry curvature and plays a central role
in nonlinear Hall responses. The BCD, analogous to a pseoudovector in two dimension lies in the plane\cite{Inti}
\begin{equation}\label{bcd}
D_{d}
=
\sum_n \int_{\mathrm{BZ}}
\frac{d^2k}{(2\pi)^2}
f_n^0(\mathbf{k})
\frac{\partial \Omega_n^c(\mathbf{k})}{\partial k_d},
\end{equation}
where d $\in$  {a,b}.

For numerical evaluation and physical interpretation, the above
expression can be conveniently rewritten using integration by parts in
the Brillouin zone\cite{WannBerri}:
\begin{equation}\label{bcd_ex}
D_{d}
=
-\sum_n \int_{\mathrm{BZ}}
\frac{d^2k}{(2\pi)^2}
\frac{\partial f_n^0(\mathbf{k})}{\partial k_d}
\Omega_n^c(\mathbf{k}),
\end{equation}
where the boundary term vanishes owing to the periodicity of the
Brillouin zone. This form directly shows that the BCD is determined by
the Berry curvature weighted by the momentum-space variation of the
occupation function.

Using
\begin{equation}
\frac{\partial f_n^0}{\partial k_d}
=
\frac{\partial f_n^0}{\partial \varepsilon_n}
\frac{\partial \varepsilon_n}{\partial k_d},
\end{equation}
the BCD can be expressed in terms of the band velocity and the energy
derivative of the Fermi distribution. In the low-temperature limit,
$-\partial f_n^0/\partial\varepsilon_n$ approaches
$\delta(\varepsilon_n-\mu)$, indicating that the BCD is predominantly
determined by the Berry curvature and band velocity in the vicinity of
the Fermi surface. Consequently, a large BCD can emerge when the Fermi
surface intersects regions of strongly enhanced and asymmetric Berry
curvature.

The calculated BCD therefore provides a direct connection between the
electronic band structure and the nonlinear Hall response. In
particular, strong variations of Berry curvature near avoided crossings,
band inversions, or other topologically nontrivial regions can
substantially enhance the BCD. The magnitude and sign of $D_{dc}$ are
also sensitive to the chemical potential, making the nonlinear Hall
response tunable through carrier doping or external gating. In the following section, we describe how this symmetry constraint is engineered in borophosphene-derived structures through controlled $\mathcal{P}$ symmetry breaking.

\section{Crystal Structure, Stability, and Symmetry Landscape}

\begin{figure}[h]
    \centering
    \includegraphics[width=0.8\linewidth]{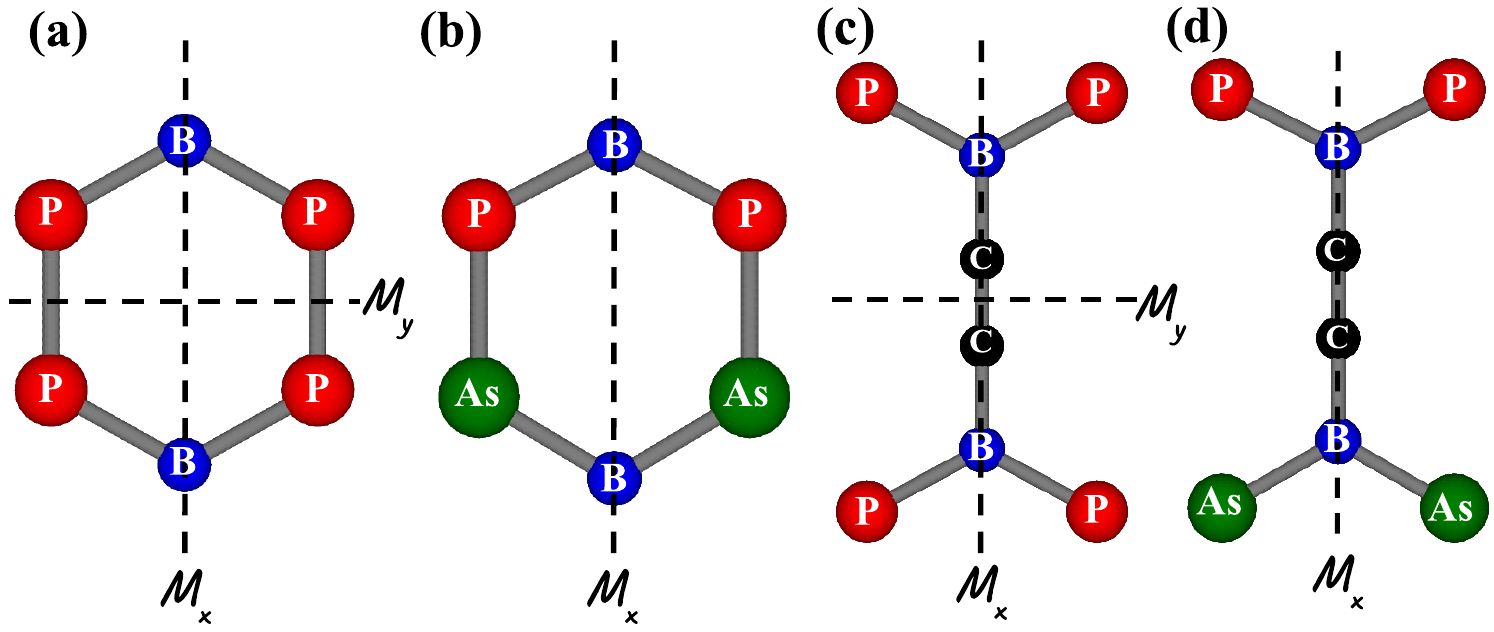}
    \captionsetup{
    justification=centerlast,
    singlelinecheck=false, 
    labelfont=bf,          
    font=small}
    \caption{Schematic representation of mirror plane symmetries and spatial inversion symmetry breaking. (a, c) Pristine MBP and AMBP monolayers, exhibiting two vertical mirror planes and a center of inversion symmetry. (b, d) Functionalized As@MBP and As@AMBP structures, where substitution of a P atom (red) with an As atom (green) breaks spatial inversion symmetry and reduces the point-group symmetry to a single vertical mirror plane.}
    \label{fig:1}
\end{figure}

\begin{sidewaystable*}[htp]
\caption{\label{tab:symmetry_progression}Systematic progression of point group symmetries, space groups, order of the point group, and inversion symmetry ($\mathcal{P}$) status across graphene, hexagonal borophosphene, and the studied 2D derivatives.}
\begin{ruledtabular}
\begin{tabular}{lccccc}
\textbf{Material} & \textbf{Point Group} & \textbf{Order} & \textbf{Inversion ($\mathcal{P}$)} & \textbf{Space Group} & \textbf{Degeneracy at the Dirac Point} \\
\hline
Graphene & $D_{6h}$ & 24 & Preserved & $P6/mmm$ (191) & Preserved (gapless) \\
Hexagonal \\Borophosphene & $D_{3h}$ & 12 & Broken & $P\bar{6}m2$ (187) & Lifted (gapped) \\
MBP / AMBP & $D_{2h}$ & 8 & Preserved & $Pmmm$ (47) & Preserved (gapless)\\
As@MBP / As@AMBP & $C_{2v}$ & 4 & Broken & $Pmm2$ (25) & Lifted (gapped) \\
\end{tabular}
\end{ruledtabular}
\end{sidewaystable*}


We study two graphene-like orthorhombic structures: modified borophosphene (MBP) and acetylene-mediated borophosphene (AMBP), as described by Zhang \textit{et al.}~\cite{MBP-main} and Subramanian \textit{et al.}~\cite{Subramanian}, respectively. MBP is derived from the hexagonal-borophosphene monolayer (space group $P\bar{6}m2$), which is a well-known semiconducting material~\cite{Borophene-review}. Later, Spatial inversion symmetry is incorporated into the crystal structure by rearranging the boron (B) and phosphorus (P) atoms, as illustrated in figure ~\ref{fig:1}(a), it formed homonuclear $\text{B–B}$ and $\text{P–P}$ dimerization. The preservation of inversion symmetry renders this phase a Dirac-like material~\cite{MBP-main}. The centrosymmetric crystal structure of AMBP, shown in figure ~\ref{fig:1}(c), represents another derived monolayer featuring acetylenic linkers ($-C\equiv C-$) incorporated between adjacent B atoms. This structural modification induces a momentum-space shift of the Dirac point along the high-symmetry $\mathrm{S}-\mathrm{Y}$ direction relative to pristine hexagonal borophosphene~\cite{Subramanian}.

In terms of point group symmetry, while graphene belongs to the $D_{6h}$ point group with 24 symmetry operations, hexagonal borophosphene falls under $D_{3h}$ with 12 symmetry operations, and MBP falls under $D_{2h}$ with 8 symmetry operations. Crucially, the $D_{2h}$ classification includes the $\mathcal{P}$ symmetry operation, which was absent in the parent $D_{3h}$ structure. The emergence of the nonlinear Hall effect demands the breaking of $\mathcal{P}$ symmetry. To achieve this, we realize non-centrosymmetric phases, As@MBP and As@AMBP, in this orthorhombic lattice by substituting phosphorus (P) with arsenic (As), as shown in figure ~\ref{fig:1}(b) and figure ~\ref{fig:1}(d). The symmetry breaking lifts the Dirac-point degeneracy, resulting in the opening of a narrow energy band gap in the electronic dispersion. Importantly, because arsenic substitution preserves non-magnetic order, time-reversal symmetry remains intact throughout these modifications, satisfying the essential symmetry condition for a BCD-driven nonlinear Hall response ($\mathcal{T}$ preserved, $\mathcal{P}$ broken).

Symmetry analysis confirms that this lower-symmetry phase belongs to the $C_{2v}$ point group, governed by the set of standard generators $\{E, C_2, \sigma_v, \sigma_v'\}$. Taking the $z$-axis perpendicular to the 2D plane, $C_2$ represents a two-fold rotation about the out-of-plane axis, while $\sigma_v(xz)$ and $\sigma_v'(yz)$ denote two orthogonal vertical mirror planes along and perpendicular to the B-B atomic linkage. Because all atoms lie strictly within a single two-dimensional plane, the monolayer also respects an effective horizontal mirror plane ($\sigma_h$). Space-group derivation using DFT relaxation combined with symmetry-finding algorithms (\texttt{spglib}/\texttt{FINDSYM\cite{spglib}}) identifies the corresponding space group as $Pmm2$ (No.~25). Here, the $\sigma_v$ mirror symmetry acts as a critical constraint governing the orientation and vanishing components of the resulting BCD. The systematic progression of point groups and crystal symmetries across these allotropes is summarized in Table~\ref{tab:symmetry_progression}. The surviving mirror is $\sigma_v'(yz)=M_x$, which maps $x\to-x$ and contains the B--B linkage, while $\sigma_v(xz)=M_y$ is broken by the As substitution. We therefore identify $\sigma_v \equiv M_x$ only in the sense of a single surviving vertical mirror, and we use $M_x$ consistently throughout. Thus, the single surviving mirror symmetry $M_x$ of the $C_{2v}$ point group explicitly determines which BCD components survive in these 2D monolayer systems, detailed discussion can be found in section V.

The optimized ground-state geometries of monolayer MBP, AMBP, As@MBP, and As@AMBP were systematically investigated to establish their structural, dynamical, and mechanical viability prior to evaluating their functional electronic properties. For the parent frameworks MBP and AMBP, phonon dispersion calculations confirm full dynamical stability, with all phonon modes remaining strictly positive throughout the Brillouin zone, consistent with earlier reports by Zhang \textit{et al.}\cite{Zhang2019} and Esakki Raj \textit{et al.}\cite{Esackraj2023}

For the functionalized monolayers, As@MBP retains complete acoustic and optical phonon stability, facilitating structural phase stability. Refer to fig. S1. Conversely, while all acoustic modes in As@AMBP remain frequency-positive, a soft optical branch characterized by an imaginary frequency persists across the Brillouin zone (Fig.~S2), signaling an intrinsic polarization instability reminiscent of soft-mode ferroelectricity in 3D perovskites.\cite{RevModPhys.73.515}

Complementing the phonon spectra, mechanical stability was evaluated using the 2D Born--Huang criteria ($C_{11} > 0$, $C_{66} > 0$, and $C_{11}C_{22} - C_{12}^2 > 0$).\cite{Born} Both functionalized structures satisfy these stability conditions. As@AMBP yields elastic constants of $C_{11} = 116.40\text{ N/m}$, $C_{22} = 182.52\text{ N/m}$, $C_{12} = 47.08\text{ N/m}$, and a soft shear modulus $C_{66} = 2.91\text{ N/m}$ ($\lambda_{1,2,3} = 2.91, 91.93, 206.98\text{ N/m}$; $C_{11}C_{22} - C_{12}^2 = 19028.5\text{ N}^2/\text{m}^2$), leading to pronounced elastic anisotropy ($Y_{\text{max}}/Y_{\text{min}} = 14.467$, $A_{\text{SU}} = 5.546$). In contrast, As@MBP demonstrates near-isotropic elastic response ($Y_{\text{max}}/Y_{\text{min}} = 1.165$, $A_{\text{SU}} = 0.010$) with $C_{11} = 173.93\text{ N/m}$, $C_{22} = 157.76\text{ N/m}$, $C_{12} = 49.42\text{ N/m}$, and $C_{66} = 50.16\text{ N/m}$. These foundational stability and anisotropy metrics confirm structural integrity (Figs.~S3 and Table~S1) while establishing a compelling basis for future investigations into phonon- and polarization-driven phenomena.

\section{Computational Details}

Our first-principles calculations were carried out within the framework of density functional theory (DFT) as implemented in the Vienna Ab Initio Simulation Package (\textsc{vasp})~\cite{vasp1,vasp2,vasp3}. The single-electron orbitals were expanded in a plane-wave basis within the Kohn-Sham scheme. Kinetic energy cutoffs of $460\text{ eV}$ (MBP and As@MBP) and $600\text{ eV}$ (AMBP and As@AMBP) were used to ensure numerical convergence. The interaction between core and valence electrons was treated using the projector augmented-wave (PAW) method~\cite{paw}, with the Perdew-Burke-Ernzerhof (PBE) functional employed for exchange-correlation interactions. Structural relaxations were performed until the total energy and Hellmann-Feynman forces converged to within $10^{-9} \text{ eV}$ and $0.001 \text{ eV/\AA}$, respectively. Spin-orbit coupling was excluded in all the self-consistent and Wannierization calculations reported here; this choice underlies the SOC-independence claim discussed in the Conclusion section.

The monolayer structures were modeled using a vacuum region of $20\text{ \AA}$ along the $z$ axis to suppress spurious interactions between periodic images. The Brillouin zone was sampled using a uniform $\Gamma$-centered $k$-point mesh of $16 \times 8 \times 1$ for MBP and As@MBP, and $32 \times 16 \times 1$ for AMBP and As@AMBP.

To facilitate an accurate investigation of Berry physics, we performed a Wannier transformation from the plane-wave representation to maximally localized Wannier functions (MLWFs)~\cite{mlwf} using the \textsc{Wannier90} package~\cite{wannier90,w90,wannier90-new}. The MLWF basis provides a real-space localized representation that is essential for capturing geometric-phase effects and topological properties. Wannierization was carried out using a frozen (inner) energy window of $[-2 , 2]\text{ eV}$ and an outer disentanglement window of $[-40, 50]\text{ eV}$ relative to the Fermi level set at 0 eV, with disentanglement performed via the standard projection-based procedure of Ref.~\cite{mlwf}. We validated the precision of the Wannier construction by comparing the electronic dispersion calculated from the MLWF-based tight-binding model against the original DFT plane-wave results. The excellent agreement between the energy bands near the Fermi level confirms that the MLWF basis successfully preserves the relevant electronic structure of the system. To compute the $k$-space integrals of the Berry curvature and derived transport properties, we employed the \textsc{WannierBerri} code~\cite{WannBerri} for efficient Wannier interpolation based on the constructed MLWFs. 



\begin{figure}[htb!]
    \centering
    \includegraphics[width=0.6\linewidth]{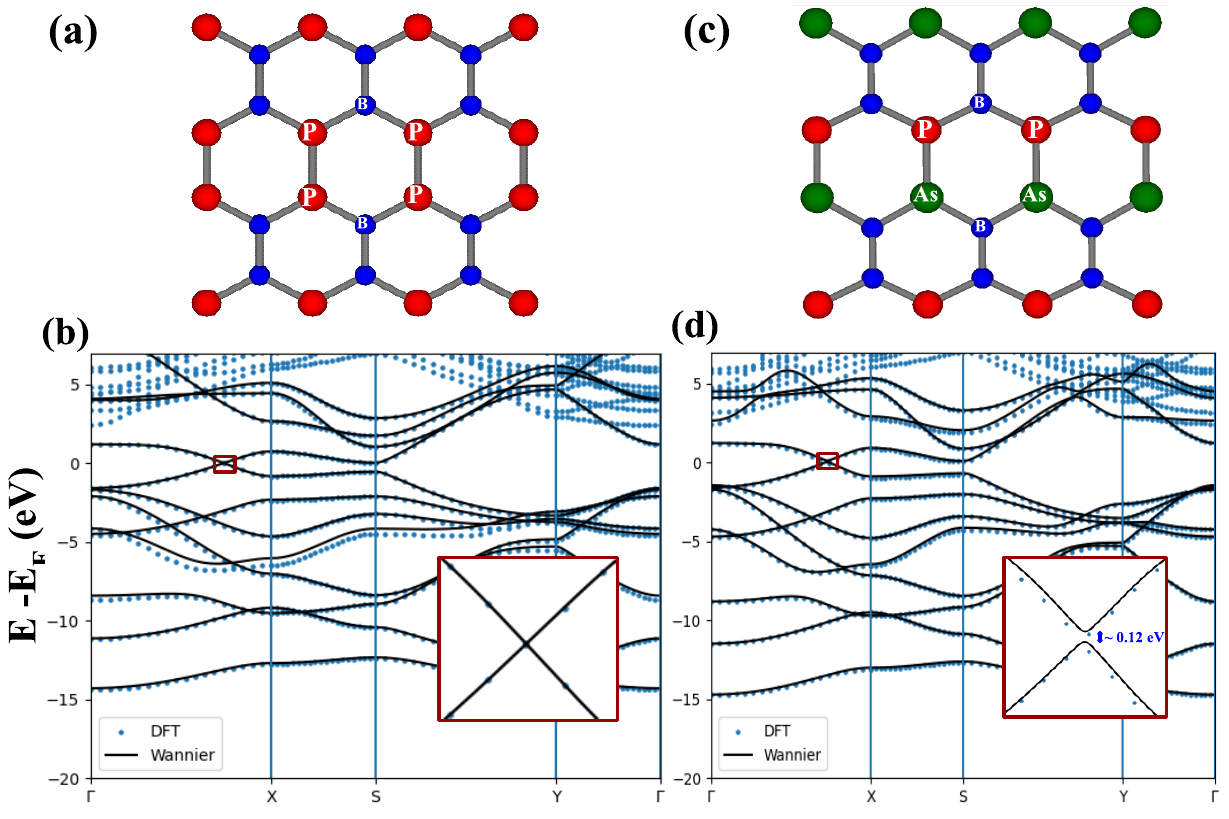}
    \captionsetup{
    justification=centerlast,
    singlelinecheck=false, 
    labelfont=bf,          
    font=small          }
    \caption{Structural and electronic properties of pristine and As-functionalized MBP. (a, c) Schematic representations of the crystal lattices for (a) centrosymmetric MBP and (b) inversion-symmetry-broken As@MBP. (b, d) Low-energy electronic band structures obtained via DFT and Wannier interpolation, illustrating the transition from (b) a gapless Dirac state to (d) a gapped phase upon As-substitution.}
    \label{fig:placeholder}
\end{figure}

\begin{figure}[h]
    \centering
    \includegraphics[width=0.6\linewidth]{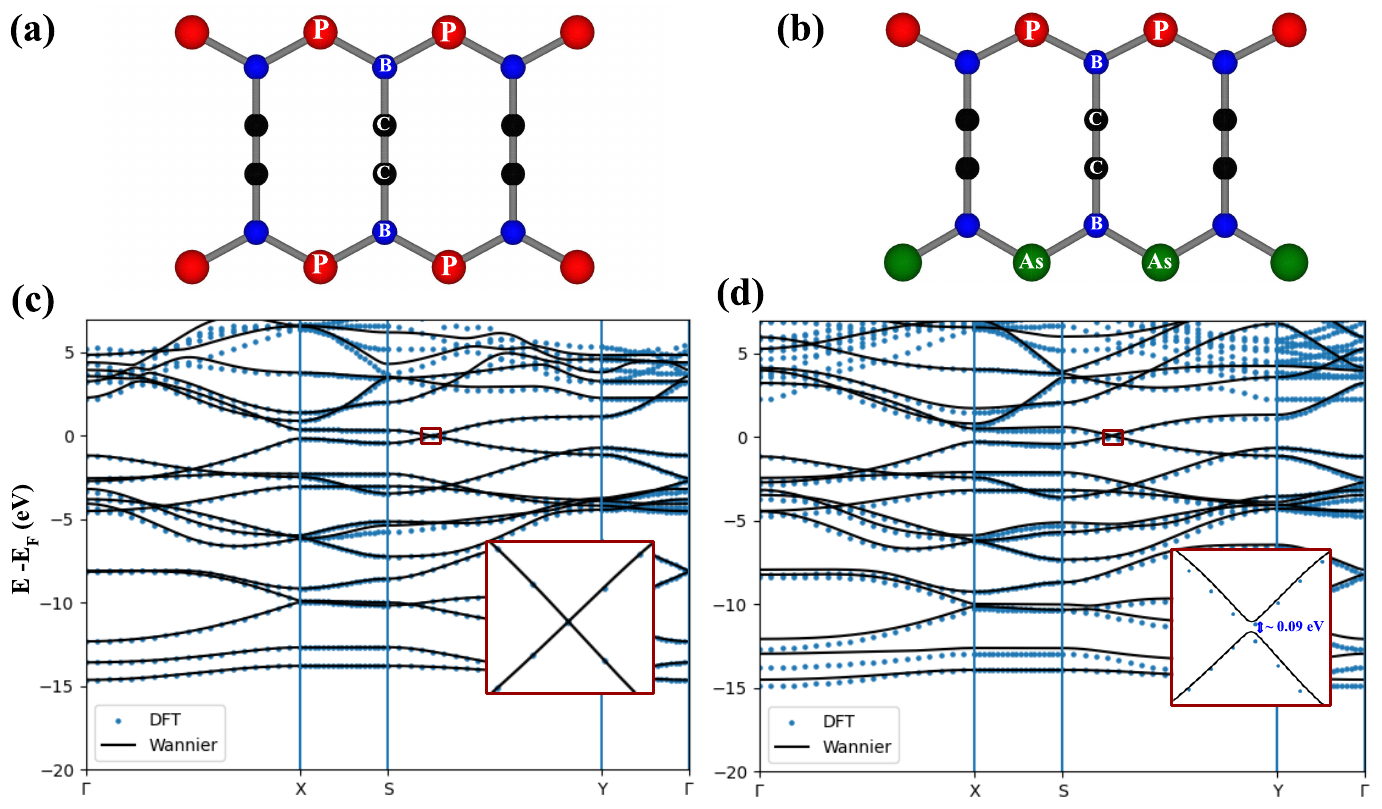}
    \captionsetup{
    justification=centerlast,
    singlelinecheck=false, 
    labelfont=bf,          
    font=small          }
    \caption{Structural and electronic properties of pristine and As-functionalized AMBP. (a, b) Schematic representations of the crystal lattices for (a) centrosymmetric AMBP and (b) inversion-symmetry-broken As@AMBP. (c, d) Low-energy electronic band structures obtained via DFT and Wannier interpolation, illustrating the transition from (c) a gapless Dirac state to (d) a gapped phase upon As-substitution.}
    \label{fig:AMBP}
\end{figure}




In the semiclassical Boltzmann framework within the constant relaxation-time approximation, the DC rectification part of the second-order nonlinear Hall conductivity is given by $\chi =\dfrac{e^{3}\tau}{\hbar^{2}}D$~\cite{Binghai,TanusriSDG,Yan}, where $D$ is the BCD defined in Eq.~\eqref{bcd_ex}, $e$ is the electron charge, and $\tau$ is the momentum relaxation time. We take $\tau \sim 1$~ps as a representative value for two-dimensional materials. Since $\chi \propto \tau$, the results can be rescaled linearly for other values of $\tau$. This approach captures only the intrinsic BCD contribution and neglects extrinsic mechanisms such as skew scattering and side jump.\\


\section{Results}

\subsection{Electronic band structure}

We started with the MBP structure shown in figure \ref{fig:placeholder} (a), which represents a recent discovery in the field of 2D Dirac materials \cite{MBP-main}. The linear electronic dispersion near the $k$-point along the $\Gamma-X$ path renders it highly promising for high-performance electronic devices, as described by the authors in Ref. \cite{MBP-main}. In our research, we reproduced the electronic dispersion for MBP shown in figure \ref{fig:placeholder} (b) using both the plane-wave Kohn-Sham density functional theory (DFT) basis and the real-space maximally localized Wannier function (MLWF) basis. The DFT- and Wannier- interpolated band structures agree remarkably well across the entire energy range. 

The pristine system is symmetric under $\mathcal{T}$ and $\mathcal{P}$, which forces the Berry curvature to vanish globally via $\mathcal{T}\Omega_{n}^c(\mathbf{k}) = -\Omega_{n}^c(-\mathbf{k})$ and $\mathcal{P}\Omega_{n}^c(\mathbf{k}) = \Omega_{n}^c(-\mathbf{k})$. Transport phenomena driven by the intrinsic Berry curvature of the material remain inactive (zero net contribution) until one of these fundamental symmetries is broken. For instance, in the anomalous Hall effect, $\mathcal{T}$ must be broken to obtain a net contribution from the Berry curvature. 

To observe the non-linear Hall effect, we break inversion symmetry by replacing one P atom with an As atom, yielding the As@MBP structure shown in figure \ref{fig:placeholder} (c). The breaking of $\mathcal{P}$ lifts the Dirac point degeneracy, opening a small energy gap of approximately 0.12 eV, while leaving the remainder of the electronic band structure virtually unchanged.

Next, we simulated another Dirac-like material, AMBP-which, like MBP, is derived from hexagonal borophosphene-that has been identified as a promising candidate cathode material for Li-ion battery applications by Subramanian \textit{et al}. \cite{Subramanian}. The centrosymmetric structure of AMBP is shown in figure \ref{fig:AMBP} (a), and its corresponding electronic structure is shown in figure \ref{fig:AMBP} (b). In an analogous fashion, breaking $\mathcal{P}$ yields the non-centrosymmetric As@AMBP structure shown in figure \ref{fig:AMBP} (c). As illustrated in figure \ref{fig:AMBP} (d), the DFT and Wannier-interpolated electronic band structures reveal a small energy gap opening of approximately 0.09 eV, with the surrounding band topology preserved. 

The slightly smaller gap induced in As@AMBP (0.09 eV) compared to As@MBP (0.12 eV) stems from structural variations in bond lengths and local strain as shown in Tables \ref{tab:mbp_vs_as_mbp} and \ref{tab:ambp_vs_as_ambp}, $As$ substitution leads to a systematic changes of both the bond lengths by $\sim$ 2\% and bond angles by $\sim$ 3$^\circ$ . Notably, both opened gaps (90–120 meV) significantly exceed the room-temperature thermal energy ($k_B T \approx 25\text{ meV}$), which suppresses unwanted thermal excitation at room temperature and confirms their viability for high-performance device applications. A detailed analysis of how this broken $\mathcal{P}$ generates a non-zero BCD in these systems is provided in Section V.B.

\begin{table}[h]
  \centering
  \caption{Structural Bond Parameters: MBP vs. As@MBP (\small\textit{Note: There is a slight increment in bond length upon As incorporation.})}
  \label{tab:mbp_vs_as_mbp}
  \begin{tabular}{lc lc}
    \toprule
    \multicolumn{2}{c}{\textbf{MBP}} & \multicolumn{2}{c}{\textbf{As@MBP}} \\
    \cmidrule(lr){1-2} \cmidrule(lr){3-4}
    Parameter & Value & Parameter & Value \\
    \midrule
    \multicolumn{4}{l}{\textit{Bond Lengths}} \\
    $P_1 - P_2$ & 2.10\,\AA & $P - \text{As}$ & 2.15\,\AA \\
    $P_2 - B_1$ & 1.84\,\AA & $\text{As} - B_1$ & 1.88\,\AA \\
    \addlinespace
    \multicolumn{4}{l}{\textit{Bond Angles}} \\
    $\angle B_2 - P_1 - P_2$ & $119.26^\circ$ & $\angle P - \text{As} - B_1$ & $121.08^\circ$ \\
    \bottomrule
  \end{tabular}
  \\[4pt]

\end{table}


\begin{table}[h]
  \centering
  \caption{Structural Bond Parameters: AMBP vs. As@AMBP}
  \label{tab:ambp_vs_as_ambp}
  \begin{tabular}{lc lc}
    \toprule
    \multicolumn{2}{c}{\textbf{AMBP}} & \multicolumn{2}{c}{\textbf{As@AMBP}} \\
    \cmidrule(lr){1-2} \cmidrule(lr){3-4}
    Parameter & Value & Parameter & Value \\
    \midrule
    \multicolumn{4}{l}{\textit{Bond Lengths}} \\
    $P_1 - P_2$ & 2.10\,\AA & $\text{As}-P$ & 2.16\,\AA \\
    $P_2 - B_1$ & 1.86\,\AA & $P - B_1$ & 1.83\,\AA \\
    $P_1 - B_2$ & 1.86\,\AA & $B_2 - \text{As}$ & 1.88\,\AA \\
    \addlinespace
    \multicolumn{4}{l}{\textit{Bond Angles}} \\
    $\angle P_2 - B_1 - P_2$ & $123.23^\circ$ & $\angle P - B_1 - P$ & $120.00^\circ$ \\
    $\angle P_1 - B_2 - P_1$ & $120.00^\circ$ & $\angle P_1 - B_2 - P_1$ & $123.00^\circ$ \\
    \bottomrule
  \end{tabular}
\end{table}

\subsection{Geometric Encoding in the electronic band structure}

\begin{figure*}
   \centering
    \includegraphics[width=\linewidth]{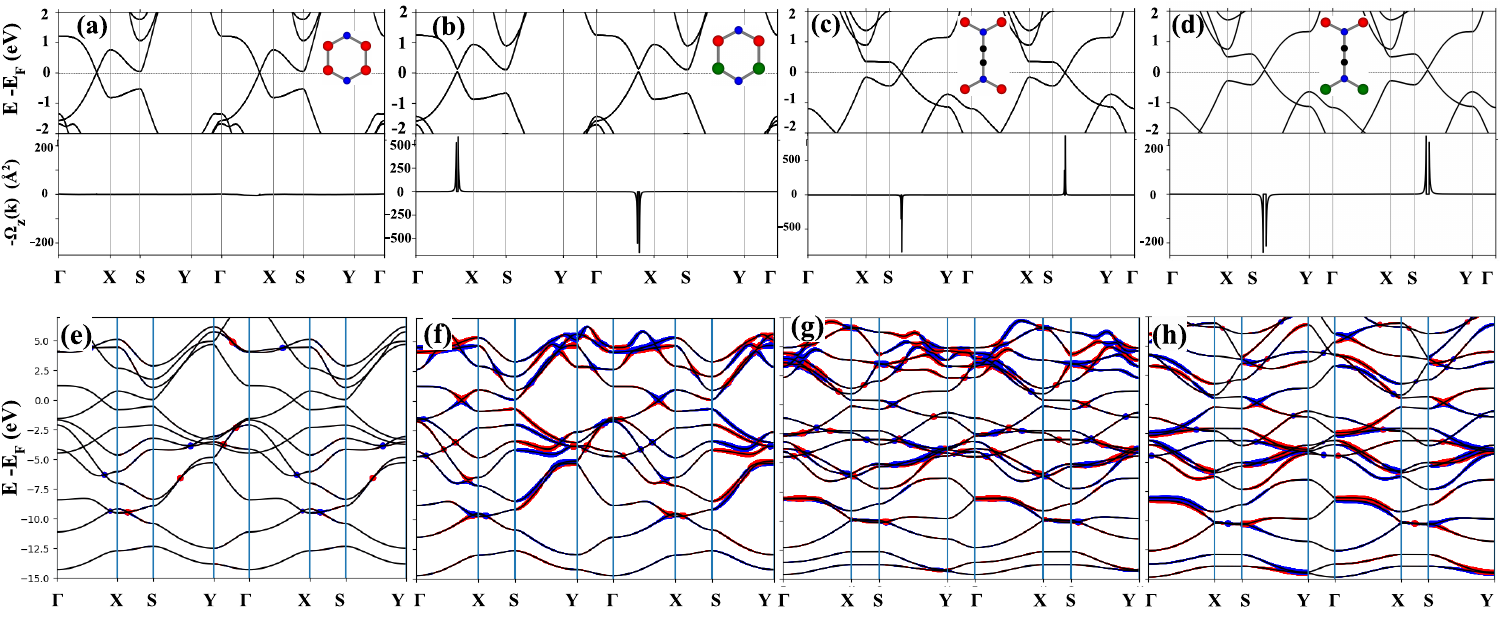}
    \captionsetup{
    justification=centerlast,
    singlelinecheck=false, 
    labelfont=bf,          
    font=small          }
    \caption{Geometric-Berry curvature in the electronic band structures for the materials. (a–d) Low-energy Wannier-interpolated band structures overlaid with the total Berry curvature in $k$-space. (e–h) Corresponding momentum-resolved Berry curvature contributions for individual bands within the first Brillouin zone. Berry curvature distribution is color-coded with red and blue representing positive and negative values, respectively. Panels are organized for comparison: MBP (a, e), As@MBP (b, f), AMBP (c, g), and As@AMBP (d, h).}
    \label{fig:poster1}
\end{figure*}

In the many-body electron system, the geometric phase information is encoded within the electronic band structure in terms of the Berry curvature\cite{TopBanTh}. This quantity is central to the emergence of global topological invariants (such as Chern number, $Z_2$ invariant) and local geometric properties within $k$-space\cite{Haldane-Nobel,Zoo:Topo}. Our calculations demonstrate the emergence and distribution of the Berry curvature in the electronic band structure in the presence of symmetry in the system.  The total Berry curvature is highly localized, specific to some k-point in the Brillouin zone\cite{BerPhaseElecPro,Berry} shown in figure \ref{fig:poster1}. Because of the broken $\mathcal{P}$ symmetry of the system, the Berry curvature exhibits an asymmetric distribution and maintains opposite signs at k-points related by $\mathcal{T}$ symmetry. This asymmetric distribution and the resulting non-zero BCD, defined as the gradient of the Berry curvature in k-space, facilitate the emergence of the nonlinear Hall effect\cite{TaIrTe4}. Consequently, the transport properties of the material are governed by these underlying geometric gradients\cite{Inti,qGeo-origin}.

In two dimensional system, the Berry curvature $\Omega_c(\mathbf{k})$, non-zero component is perpendicular to the $ab$-plane and behaves like a pseudoscalar quantity. The vertical mirror plane $\sigma_v $ specifically $M_a$ acts as a mirror reflection on the momentum coordinate as $(k_a, k_b) \to (-k_a, k_b)$. The Berry curvature under the reflection $M_a$ changes sign:
$$\Omega_c(k_a, k_b) = -\Omega_c(-k_a, k_b)$$

Under the reflection $\mathcal{M}_a$ across the mirror line at $k_a = 0$, the out-of-plane Berry curvature $\Omega_c(\mathbf{k})$ transforms as an odd function with respect to $k_a$, satisfying the above relation, which implies that its derivative $\partial \Omega_c / \partial k_a$ is even in $k_a$, whereas $\partial \Omega_c / \partial k_b$ remains odd in $k_a$. As defined, the BCD   transforms as an in-plane pseudovector \cite{Inti,Yan}, and its total value is governed by the global symmetry of the Brillouin zone: the integrand for the $b$-component, $D_b \propto \int f_0(\mathbf{k}) (\partial \Omega_c / \partial k_b) \, d^2k$, is odd in $k_a$, causing contributions from opposite sides of the mirror line to cancel identically and forcing $D_b = 0$, whereas the integrand for the $a$-component, $D_a \propto \int f_0(\mathbf{k}) (\partial \Omega_c / \partial k_a) \, d^2k$, is even in $k_a$ and yields a non-vanishing net value when the distribution is asymmetric along the $a$-direction. Consequently, while a horizontal mirror plane $\sigma_h$ does not constrain $D_a$ or $D_b$, a vertical mirror plane $\sigma_v$ ($\mathcal{M}_a$) strictly forces $D_b$ to vanish and restricts the dipole vector to point along the $a$-axis; as a result, the nonlinear Hall current $J_a = \chi_{abc} E_b E_c$, mediated by the BCD-proportional susceptibility tensor $\chi$, generates a transverse nonlinear response upon applying an electric field $E_a$ along the $a$-direction that directly probes the Berry curvature distribution around the $\mathcal{M}_a$ mirror line\cite{Binghai}.

In figure \ref{fig:poster1} we show the electronic band structure after the Wannier interpolation and the net Berry curvature in the BZ for our systems. MBP is the Dirac material with the preservation of $\mathcal{T}$ and $\mathcal{P}$, and as a consequence the Berry curvature vanishes everywhere in the k-space, shown in figure \ref{fig:poster1} (a). The Berry curvature becomes finite once the system becomes gapped by incorporation of breaking of $\mathcal{P}$ in the system As@MBP figure \ref{fig:poster1} (b). In the similar way, for AMBP the Berry curvature is zero everywhere except it diverges only at the k-points where the energies become Dirac degenerate. And for As@AMBP gives finite Berry curvature, as it is a broken inversion-symmetric material. 

Figures \ref{fig:poster1} (e,f,g,h) show the band-resolved, momentum-dependent Berry curvature along the $\Gamma$–X–S–Y–$\Gamma$–X–S–Y path, with red and blue markers denoting positive and negative contributions, respectively. 
It is a visual on how the symmetry of the system changes the geometric quantity in the electronic band structure and the corresponding k-point with respect to the energy is contributed. One can compare the figures how the Berry curvature gets concentrated at the band curve once the degeneracy is  lifted due to symmetry breaking mechanism. In all four systems the Berry curvature is negligible over most of the Brillouin zone and is strongly localized at isolated $k$-points where bands approach or cross, consistent with its inverse-square dependence on the interband energy gap. In MBP (e) these hot spots are sparse and point-like, with alternating signs at neighboring Dirac-like nodes, reflecting opposite Berry phases that partially cancel in integrated responses. In the As-containing MBP (refer figure 4f), the increased band entanglement produces more numerous and extended hot spots of both signs, particularly at $-10$ to $-2.5$~eV and above $E_F$. On the other hand, As@AMBP (h) exhibits a markedly larger number of band anticrossings, and the Berry curvature becomes more extended along the band dispersions, with pronounced positive and negative lobes of both signs. Thus, the intrinsic Berry-curvature response is therefore governed by the curvature near $E_F$ at the small gaps that regularize the Dirac points.

\begin{figure}[htp!]
    \centering
    \includegraphics[width=0.6\linewidth]{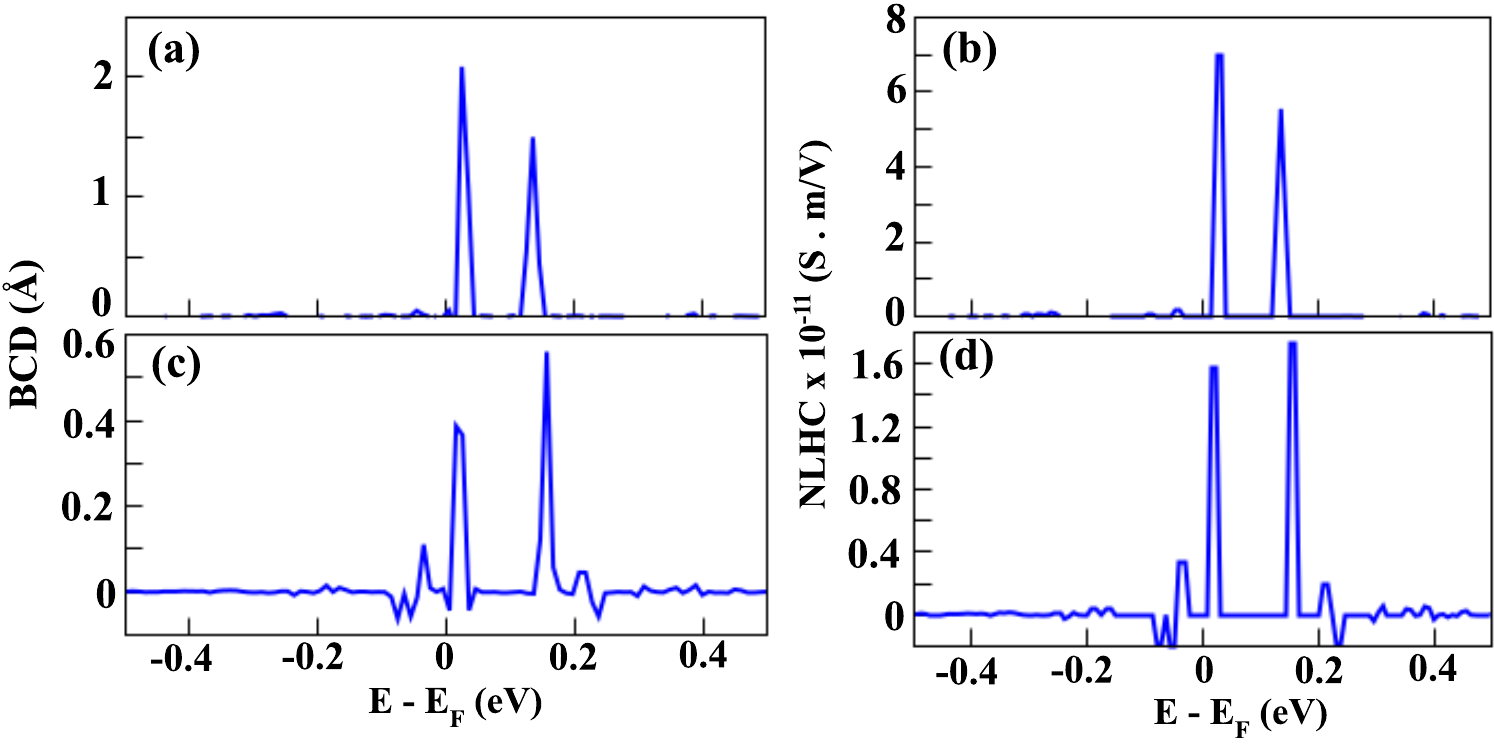}
    \captionsetup{
    justification=centerlast,
    singlelinecheck=false, 
    labelfont=bf,          
    font=small}
    \caption{Nonlinear transport properties for As@MBP and As@AMBP. (a, c) BCD and (b, d) corresponding nonlinear Hall conductivity as a function of Energy. Top panel (a–b) represent the As@MBP system, bottom panel (c–d) represent As@AMBP.}
    \label{fig:bcdnlahc}
\end{figure}

Further, we calculated the BCD, $D_a(E)$, and the corresponding second-order nonlinear Hall conductivity as a function of energy relative to the Fermi level, as shown in Figure~\ref{fig:bcdnlahc}. The BCD is shown in Figures~\ref{fig:bcdnlahc}(a) and \ref{fig:bcdnlahc}(c) for As@MBP and As@AMBP, respectively, while the corresponding nonlinear Hall conductivity is shown in Figures.~\ref{fig:bcdnlahc}(b) and \ref{fig:bcdnlahc}(d). The conductivity arises purely from the intrinsic BCD of the material~\cite{WannBerri,Inti,Binghai}.
For As@MBP, the BCD exhibits two sharp,
well-separated peaks, reaching $D_a \approx 2.1$~\AA\ at
$E-E_F\approx0.02$~eV and $D_a\approx1.5$~\AA\ at
$E-E_F\approx0.15$~eV, with the corresponding NLHC peaking at
$\approx7.0\times10^{-11}$~S$\cdot$m/V and $\approx5.6\times10^{-11}$~S$\cdot$m/V,
respectively. Both quantities remain vanishingly small outside this
narrow $\sim0.2$~eV energy window, consistent with the strong
localization of the underlying Berry curvature to specific bands and
$k$-points near $E_F$ discussed above.

As@AMBP displays a qualitatively richer energy dependence: in addition
to two dominant positive peaks, $D_a\approx0.38$~\AA\ and
$D_a\approx0.58$~\AA, the BCD changes sign twice within the same
energy window, with a small negative excursion just below $E_F$ and a
second, more pronounced negative dip near $E-E_F\approx0.2$--$0.25$~eV;
the corresponding NLHC tracks this sign structure closely. This
demonstrates explicitly that the sign of the nonlinear Hall response in
As@AMBP is tunable through the Fermi level alone (via, e.g., gating or
doping), whereas no such sign reversal is observed for As@MBP over the
same energy range.

Comparing the two materials, the peak BCD of As@MBP
($\approx2.1$~\AA) exceeds that of As@AMBP ($\approx0.58$~\AA) by a
factor of $\sim3.6$, and the peak NLHC is correspondingly $\sim4\times$
larger.
%
%
We note that the relative NLHC peak heights in As@AMBP do not scale
linearly with the corresponding BCD peak heights (e.g., the second BCD
peak is only moderately larger than the first, while the two NLHC peaks
are nearly equal); this reflects the additional dependence of the
nonlinear Hall conductivity on the band velocity and relaxation time
entering the full transport expression, beyond the BCD magnitude alone.

\subsection{Model Hamiltonian}

 For the MBP system, the electronic structure exhibits distinct anisotropy. Along the $\Gamma-X$ high-symmetry path, corresponding to the $k_x$ direction in reciprocal space, the valence and conduction bands intersect linearly at the Fermi level ($E_F$). This linear dispersion characterizes a zero-gap Dirac cone centered at a specific $k$-point on this axis. Conversely, the material maintains a robust, wide band gap along all other directions in the Brillouin zone, suggesting that the relativistic-like carrier physics is confined to a specific crystallographic orientation.

In contrast, in the AMBP system, the Dirac degeneracy is shown along the $S-Y$ direction corresponding to the $k_y$ direction in reciprocal space and the electronic dispersion becomes flipped compared to the MBP system. These findings demonstrate that by preserving specific crystalline symmetries, it is possible to realize and manipulate Dirac fermions. Furthermore, the position of the Dirac degeneracy is not fixed but can be systematically tuned between different $k$-points within the Brillouin zone by altering the material's structural or chemical environment. 

Upon targeted substitution with an arsenic ($As$) atom, the structural symmetry of both precursors is reduced, resulting in a non-centrosymmetric pair: As@MBP and As@AMBP. This explicit breaking of $\mathcal{P}$ classifies these systems as promising candidates for the observation of the nonlinear Hall effect (NLHE), which typically requires a non-vanishing BCD. 

\begin{figure}[htp!]
   \centering
  \includegraphics[width=0.4\linewidth]{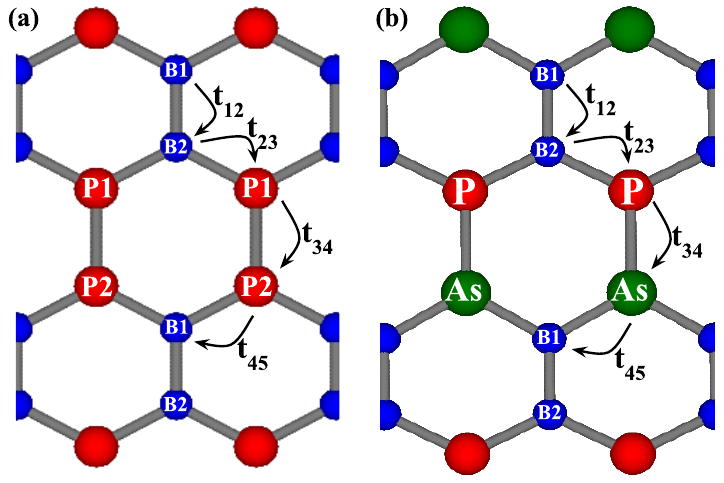}
  \captionsetup{
    justification=centerlast,
    singlelinecheck=false, 
    labelfont=bf,          
    font=small          }
    \caption{Schematic of the tight-binding lattice models for \textbf{(a)} MBP and \textbf{(b)} As@MBP. The diagrams illustrate the atomic arrangement and the relevant hopping parameters considered in the TB Hamiltonian.}
   \label{fig:lattice_MBP}
\end{figure}

\begin{figure}[htp!]
   \centering
  \includegraphics[width=0.4\linewidth]{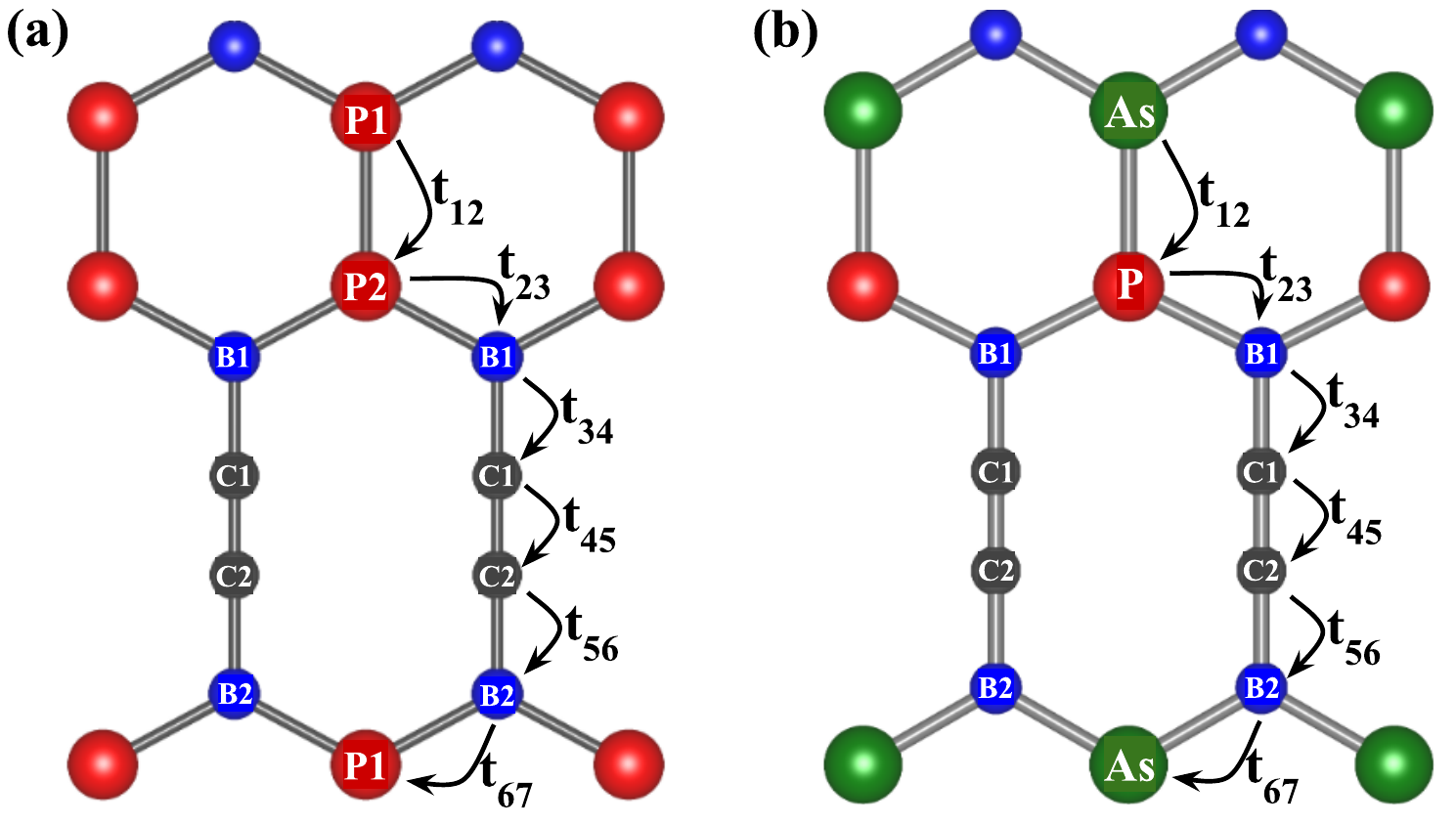}
  \captionsetup{
    justification=centerlast,
    singlelinecheck=false, 
    labelfont=bf,          
    font=small          }
    \caption{Schematic of the tight-binding lattice models for \textbf{(a)} AMBP and \textbf{(b)} As@AMBP. The diagrams illustrate the atomic arrangement and the relevant hopping parameters considered in the TB Hamiltonian.}
   \label{fig:lattice_AMBP}
\end{figure}

Consequently, these phases of the study focus on a tight-binding analysis to quantify the nearest-neighbor hopping integrals between distinct atomic sites. By examining these parameters, we aim to elucidate how chemical substitution modifies the electronic hopping amplitudes of the system. To describe the dispersion relations of these two-dimensional systems, we formulated a tight-binding (TB) Hamiltonian based on the localized atomic orbitals using MLWF \cite{wannier90}and it nicely predicts the low energy electronic states,

The tight-binding Hamiltonian shown in Eq.~(14) is given by
\begin{equation}
\mathcal{H} = \sum_{i,\alpha} \epsilon_{i,\alpha} c_{i,\alpha}^\dagger c_{i,\alpha} + \sum_{i < j} \sum_{\alpha, \beta} \left( t_{ij}^{\alpha\beta} c_{i,\alpha}^\dagger c_{j,\beta} + \text{H.c.} \right),
\end{equation}
where $\mathcal{H}$ represents the total Hamiltonian operator of the system. The first term describes the on-site potential, where $\epsilon_{i,\alpha}$ is the on-site energy of an electron occupying orbital $\alpha$ at atomic site $i$, $c_{i,\alpha}^\dagger$ ($c_{i,\alpha}$) is the creation (annihilation) operator for an electron at site $i$ with orbital index $\alpha$, and $c_{i,\alpha}^\dagger c_{i,\alpha} = \hat{n}_{i,\alpha}$ is the particle number operator. 

The second term describes the kinetic hopping process between different atomic sites and orbitals. Here, $t_{ij}^{\alpha\beta}$ is the hopping matrix element between orbital $\beta$ on site $j$ and orbital $\alpha$ on site $i$. The summation $\sum_{i < j}$ restricts the site index pairs to unique bonds to prevent double-counting upon adding the Hermitian conjugate ($\text{H.c.}$), which accounts for the reverse hopping process ($j \rightarrow i$) and ensures that the Hamiltonian remains Hermitian.


The nearest-neighbor (NN) hopping  amplitude between adjacent atomic sites is illustrated in figure \ref{fig:lattice_MBP} for MBP and As@MBP and in figure \ref{fig:lattice_AMBP} for AMBP and As@AMBP. The corresponding changes in the hopping energy are mentioned in Tables \ref{tab:mbp_hopping} and \ref{tab:ambp_hopping} respectively. During the study of the hopping energy we consider the $p_z-p_z$ orbital overlap between the adjacent NN atoms.

\begin{figure}
    \centering
    \includegraphics[width=0.25\textwidth]{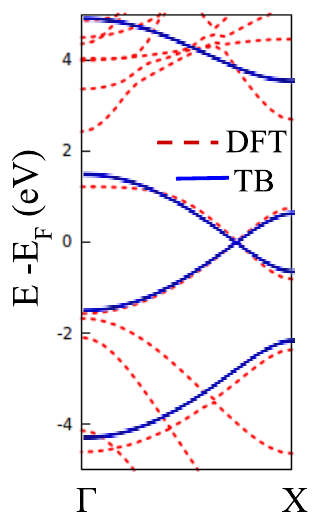}
    \captionsetup{
    justification=centerlast,
    singlelinecheck=false, 
    labelfont=bf,          
    font=small          }
    \caption{Calculated band structure using wannier tight-binding (TB) model (blue lines) and DFT (red dotted lines) along $\Gamma(0, 0, 0)$ to $X(0.5, 0, 0)$ for MBP}
\label{tbBand}
\end{figure}

Figure~\ref{tbBand} compares the DFT band structure of MBP with
that obtained from the Wannier-interpolated tight-binding (TB) Hamiltonian
along $\Gamma(0,0,0) \to X(0.5,0,0)$. The TB bands reproduce the DFT
dispersion almost exactly over the full energy window shown near $E_F$,
confirming that the Wannierization procedure faithfully captures the
low-energy electronic structure relevant to the topological quantities
discussed in Sec.~V. In particular, the two valence and two conduction
bands cross near $E-E_F \approx 0$ at a generic point along the path,
consistent with the Dirac-semimetal character of MBP protected by the
combination of $\mathcal{T}$ and $\mathcal{P}$ symmetry discussed
above. Additional DFT bands appearing above $\sim+2$~eV and below
$\sim-2$~eV are not reproduced by the TB model, as the Wannier functions
were constructed from a restricted energy window chosen to isolate the
bands nearest $E_F$ and this is in agreement with the earlier study made by Zhang \textit{et
al.} \cite{MBP-main}. These higher-energy states do not contribute
appreciably to the Berry curvature or BCD in the doping range considered
in this work.
\begin{sidewaystable*}[htbp]
  \centering
  \caption{Calculated hopping integrals $t_{ij}$ (in eV) between $\mathrm{p}_z$-orbitals for the MBP and As@MBP systems.}
  \label{tab:mbp_hopping}
  \vspace{4pt}
  \begin{tabular}{l c c c c}
    \toprule
    & \multicolumn{2}{c}{\textbf{MBP}} & \multicolumn{2}{c}{\textbf{As@MBP}} \\
    \cmidrule(lr){2-3} \cmidrule(lr){4-5}
    \textbf{Hopping Term} ($t_{ij}^{\alpha\beta}$) & \textbf{Bond Length} (\AA) & \textbf{Hopping energy} (eV) & \textbf{Bond Length} (\AA) & \textbf{Hopping energy} (eV) \\
    \midrule
    $t_{12}^{\mathrm{p}_z-\mathrm{p}_z}$ & $\text{B}_1-\text{B}_2$ ($1.65$) & $-1.81$ & $\text{B}_1-\text{B}_2$ ($1.60$) & $-0.91$ \\
    $t_{23}^{\mathrm{p}_z-\mathrm{p}_z}$ & $\text{B}_2-\text{P}_1$ ($1.84$) & $-1.33$ & $\text{B}_2-\text{P}$ ($1.81$) & $-1.23$ \\
    $t_{34}^{\mathrm{p}_z-\mathrm{p}_z}$ & $\text{P}_1-\text{P}_2$ ($2.10$) & $-0.05$ & $\text{P}-\text{As}$ ($2.15$) & $-0.03$ \\
    $t_{45}^{\mathrm{p}_z-\mathrm{p}_z}$ & $\text{P}_2-\text{B}_1$ ($1.84$) & $-1.33$ & $\text{As}-\text{B}_1$ ($1.88$) & $-1.37$ \\
    \bottomrule
  \end{tabular}
\end{sidewaystable*}

\begin{sidewaystable*}[htbp]
  \centering
  \caption{Calculated hopping integrals $t_{ij}$ (in eV) between $\mathrm{p}_z$-orbitals for the AMBP and As@AMBP systems.}
  \label{tab:ambp_hopping}
  \vspace{4pt}
  \begin{tabular}{l c c c c}
    \toprule
    & \multicolumn{2}{c}{\textbf{AMBP}} & \multicolumn{2}{c}{\textbf{As@AMBP}} \\
    \cmidrule(lr){2-3} \cmidrule(lr){4-5}
    \textbf{Hopping Term} ($t_{ij}^{\alpha\beta}$) & \textbf{Bond Length} (\AA) & \textbf{Hopping energy} (eV) & \textbf{Bond Length} (\AA) & \textbf{Hopping energy} (eV) \\
    \midrule
    $t_{12}^{\mathrm{p}_z-\mathrm{p}_z}$ & $\text{P}_1-\text{P}_2$ ($2.10$) & $-3.29$ & $\text{As}-\text{P}$ ($2.16$) & $\phantom{-}1.79$ \\
    $t_{23}^{\mathrm{p}_z-\mathrm{p}_z}$ & $\text{P}_2-\text{B}_1$ ($1.86$) & $-1.37$ & $\text{P}-\text{B}_1$ ($1.83$)  & $-1.30$ \\
    $t_{34}^{\mathrm{p}_z-\mathrm{p}_z}$ & $\text{B}_1-\text{C}_1$ ($1.48$) & $-2.13$ & $\text{B}_1-\text{C}_1$ ($1.45$) & $-2.34$ \\
    $t_{45}^{\mathrm{p}_z-\mathrm{p}_z}$ & $\text{C}_1-\text{C}_2$ ($1.23$) & $-0.03$ & $\text{C}_1-\text{C}_2$ ($1.23$) & $-0.01$ \\
    $t_{56}^{\mathrm{p}_z-\mathrm{p}_z}$ & $\text{C}_2-\text{B}_2$ ($1.48$) & $-2.38$ & $\text{C}_2-\text{B}_2$ ($1.45$) & $-2.16$ \\
    $t_{67}^{\mathrm{p}_z-\mathrm{p}_z}$ & $\text{B}_2-\text{P}_1$ ($1.86$) & $-1.79$ & $\text{B}_2-\text{As}$ ($1.89$) & $\phantom{-}0.14$ \\
    \bottomrule
  \end{tabular}
\end{sidewaystable*}

%

Tables~\ref{tab:mbp_hopping} and~\ref{tab:ambp_hopping} list the calculated
$p_z$ -- $p_z$ tight-binding hopping integrals $t_{ij}^{\alpha\beta}$ for the
MBP/As@MBP and AMBP/As@AMBP ring structures, respectively, together with the
corresponding bond lengths.

In pristine MBP, the two symmetry-related B--P bonds carry identical
hopping strengths, $t_{23}=t_{45}=-1.33$~eV at equal bond length
(1.84~\AA), reflecting the mirror/inversion symmetry of the ring. Upon
incorporation of As (As@MBP), this equivalence is lifted:
$t_{23}=-1.23$~eV (B--P, 1.81~\AA) and $t_{45}=-1.37$~eV (As--B,
1.88~\AA) are no longer equal, providing a microscopic, tight-binding-level
confirmation of the inversion-symmetry breaking invoked in Sec.~V to
explain the emergence of finite Berry curvature in As@MBP. We note that
the B-B hopping $t_{12}$ decreases substantially in magnitude upon As
incorporation ($-1.81\rightarrow-0.91$~eV) despite a \emph{shorter} bond
length (1.65~$\rightarrow$~1.60~\AA), indicating that this change reflects
electronic redistribution induced by As rather than a simple
bond-length effect. All hopping integrals in As@MBP retain the same sign
as in MBP, so As substitution renormalizes hopping magnitudes without
changing the qualitative bonding character of the ring.

In contrast, the two nominally mirror-related hoppings in pristine AMBP,
$t_{23}=-1.37$~eV (P--B) and $t_{67}=-1.79$~eV (B--P), are already
inequivalent prior to As substitution.
%
%
Most notably, two hoppings in the AMBP ring undergo a full \emph{sign
reversal} upon As incorporation: $t_{12}$ changes from $-3.29$~eV
(P--P) to $+1.79$~eV (As--P), and $t_{67}$ changes from $-1.79$~eV
(B--P) to $+0.14$~eV (B--As). This is qualitatively different from the
As@MBP case, where no sign changes occur among any of the hopping
parameters. Since a sign change in the dominant hopping term reflects a
qualitative change in orbital coupling at the substitution site rather
than a simple renormalization, this is likely relevant to any
qualitative differences observed between the As@MBP and As@AMBP
Berry-curvature and BCD results reported in Sec.~V, and merits explicit
discussion.
%
%
By contrast, the C1-C2 hopping $t_{45}$ is essentially unaffected by
As substitution, remaining at the same bond length (1.23~\AA) and
near-zero magnitude ($-0.03\rightarrow-0.01$~eV) in both AMBP and
As@AMBP, consistent with its position farthest from the As substitution
site.


Figure~\ref{fig:lattice_MBP} shows the real-space bonding path underlying
the hopping integrals of Table~\ref{tab:mbp_hopping}: a four-bond vertical
ring, B1--B2--P1--P2--B1, bridging two phosphorus-rich hexagonal sheets.
Importantly, As incorporation (figure~\ref{fig:lattice_MBP} (b)) replaces
only \emph{one} of the two phosphorus sites in this ring, P2
$\rightarrow$ As, while the second phosphorus site (P1 $\rightarrow$ P)
is left intact. This single-site substitution is the concrete structural
origin of the inversion-symmetry breaking discussed in Sec.~V: it is
precisely why the two nominally equivalent B--P bonds of pristine MBP,
$t_{23}$ and $t_{45}$, become inequivalent ($t_{23}=-1.23$~eV vs.
$t_{45}=-1.37$~eV) once one of the two phosphorus neighbors is replaced
by arsenic.

Figure~\ref{fig:lattice_AMBP} shows the corresponding six-bond ring,
P1--P2--B1--C1--C2--B2--P1, for AMBP and As@AMBP
(Table~\ref{tab:ambp_hopping}), in which a carbon dimer (C1--C2) is
inserted as a linker between the two boron sites bridging the
phosphorus sheets. As in the MBP case, As substitution in
figure ~\ref{fig:lattice_AMBP}(b) occurs at a single ring site per
hexagonal unit rather than uniformly, breaking the mirror equivalence
between $t_{12}$ (P1--P2 $\rightarrow$ As--P) and $t_{67}$ (B2--P1
$\rightarrow$ B2--As) - the same two bonds identified above as
undergoing a full sign reversal upon doping. By contrast, the C1--C2
bond ($t_{45}$) lies farthest from either substitution site in the
ring, consistent with its being essentially unaffected by As
incorporation in both the bond length and the hopping energy.
%

\subsection{Strain-tunable Berry curvature and Berry curvature dipole}
\label{sec:strain}

The magnitude of the Berry curvature and of the BCD is controlled primarily by two factors: the crystalline symmetry, determines which components of the BCD are allowed to be nonzero (Sec. V (B)), and the size of the energy gap at the relevant band-touching points, which sets the overall magnitude of the response. In this section we examine how externally applied strain tunes the band gap of As@MBP and As@AMBP, and how this, in turn, enhances the Berry curvature and the BCD.

For a two-band model near opening point of the band gap, the Berry curvature scales inversely with the gap, schematically \cite{cai2013magnetic}   
\begin{equation}
\Omega(\mathbf{k}) \;\propto\; \frac{\Delta}{\left(\Delta^{2}+\hbar^{2}v^{2}k^{2}\right)^{3/2}},
\label{eq:berry_gap}
\end{equation}
where $\Delta$ is the local band gap and $v$ is the band velocity near the
touching point \cite{Joseph-awadesh}. Equation~(\ref{eq:berry_gap}) shows
explicitly that as $\Delta \to 0$ the peak Berry curvature diverges as $1/\Delta^2$, so that any mechanism which narrows the band gap while preserving the symmetry-allowed BCD component will enhance both $\Omega(\mathbf{k})$ and
$D_a$ (Eq.~(14)). On the other hand, application of strain simultaneously induces a distortion in bond length and bond angle, altering atomic orbital hybridization and modifying the interband transition matrix elements $\langle n,\mathbf{k} \vert{} \hat{v}_a \vert{} m,\mathbf{k} \rangle$ (Eq.~\ref{BC}). Thus, the resulting evolution of the Berry curvature and BCD under progressive strain is dictated by the interplay between gap reduction and orbital overlap deformation.

%
%

Applying uniaxial compressive strain along the $a$-axis of the crystal provides exactly such a mechanism; it reduces the band gap of As@MBP without breaking the symmetry that protects the nonzero BCD component. Table~\ref{tab:strain_gap} lists the band gap, computed with the PBE functional, as a function of applied compressive strain. We find that the band gap decreases monotonically with increasing compressive strain, and correspondingly the peak Berry curvature increases sharply as the gap narrows, consistent with Eq.~(\ref{eq:berry_gap}). At $4\%$ compressive strain, the band gap of As@MBP is reduced to $57$~meV, at which point the Berry curvature is markedly enhanced (refer to figure S1 from supplementary information) and decreased with further increment in strain.


%
%


\begin{table}[htbp]
\centering
\caption{Calculated band gap ($E_{\text{g}}$) of As@MBP and As@AMBP as a function of applied uniaxial compressive strain along the $a$-axis.}
\label{tab:strain_gap}
\begin{tabular}{ccc}
\toprule
\textbf{Strain} & \multicolumn{2}{c}{\textbf{As@MBP}} \\
\cmidrule(lr){2-3}
\textbf{(\%)} & \textbf{$E_{\text{g}}$ (meV)} \\
\midrule

0 & 120  \\
2 & 66   \\
4 & 57   \\
6 & 49   \\
8 & 38   \\
\bottomrule
\end{tabular}
\end{table}

\begin{figure}[htp!]
    \centering
    \includegraphics[width=0.6\linewidth]{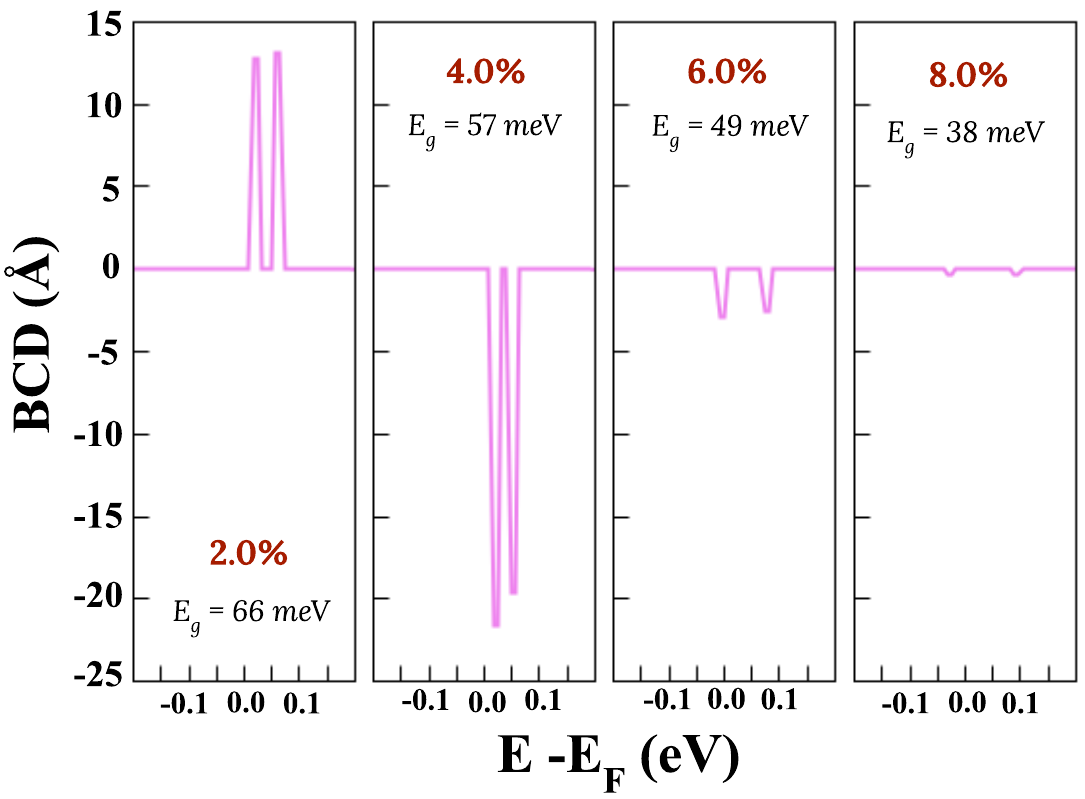}
    \captionsetup{
    justification=centerlast,
    singlelinecheck=false, 
    labelfont=bf,          
    font=small          }
    \caption{Variation of the BCD in As@MBP under uniaxial compressive strain. The plots illustrate the dependence of BCD components on the lattice parameter along the $x$-axis, highlighting the simultaneous reduction in the band gap as strain increases}
    \label{fig:strain}
\end{figure}

Figure~\ref{fig:strain} illustrates the evolution of the Berry curvature dipole (BCD), $D_a(\varepsilon)$, under uniaxial compressive strain for As@MBP. As the compressive strain increases from $2\%$ to $8\%$, the band gap narrows systematically in a monotonic manner from $66$ to $38$ meV. In contrast to this monotonic gap reduction, $D_a(\varepsilon)$ evolves non-monotonically, exhibiting redistribution and sign reversals in the vicinity of the Fermi level. Specifically, $D_a$ displays prominent positive features at $2\%$ strain, reverses to dominant negative features at $4\%$ strain which is a significant ten-fold increase in magnitude compared to the unstrained structure, and becomes substantially suppressed at higher strains. Because the second-order nonlinear Hall current is directly proportional to BCD Eq.(9), these results demonstrate that uniaxial compressive strain provides a robust mechanical knob to dynamically control over large range of magnitude with directional polarity and energy dependence of the nonlinear Hall response in As@MBP.

In contrast, As@AMBP shows a more complex strain dependence, with alternating positive and negative BCD features and pronounced peaks that persist across the investigated strain range (see Figure S4 of the Supplemental Material). Despite the reduction in the band gap, the magnitude of the BCD in As@AMBP remains comparatively robust, with no significant overall change under strain. Thus, compressive strain provides an effective tuning parameter for the nonlinear Hall response of As@MBP, whereas the relatively stable BCD of As@AMBP suggests a nonlinear Hall response that is more robust against mechanical deformation.

\section{Conclusion}


  In summary, we investigated the electronic structure, Berry curvature, and its associated dipole contribution to observe the nonlinear Hall response of two As-substituted borophosphene derivatives, As@MBP and As@AMBP. The study established a direct connection between As-induced inversion symmetry breaking and microscopic orbital coupling by opening the energy gaps of $0.12$~eV in As@MBP and $0.09$~eV in As@AMBP, above room-temperature thermal fluctuations. Both systems exhibit a finite BCD of  $2.1$ \AA  and $0.58$ \AA, respectively, without spin-orbit coupling, demonstrating that the response is primarily governed by orbital physics and symmetry breaking. 

   Tight-binding analysis of the $p_z$ hopping integrals (Tables~IV and V) shows that As incorporation affects the two host lattices in qualitatively different ways. In As@MBP, all hopping signs are preserved, but the symmetry-related B--P hoppings, equal to $-1.33$~eV in MBP, split into $-1.23$ and $-1.37$~eV, and the intradimer B--B hopping is suppressed by about 50\% (from $-1.81$ to $-0.91$~eV). The inversion-symmetry breaking is therefore carried by a quantitative renormalization of the hoppings. In contrast, As@AMBP undergoes a qualitative reconstruction of the orbital coupling at the substitution site: the As--P hopping reverses sign ($-3.29 \rightarrow +1.79$~eV), and the B--As hopping collapses from $-1.79$ to $+0.14$~eV. The remaining hoppings change by only $\lesssim 0.2$~eV, which suggests a perturbation localized around the As atom. These distinct mechanisms arise despite small structural distortions in both systems, with bond-length changes of at most $\sim0.06$~\AA\ ($\lesssim 3\%$). The nonlinear transport response is therefore governed primarily by the local orbital environment of the substitution site rather than by lattice distortion.\\
   
Our analysis of strain engineering further shows that lattice deformation changes geometric encoding in an effective means of enhancing the BCD magnitude for As@MBP. We find that strain, systematically reduced the band gap of the material and the interplay between gap reduction and orbital overlap deformation due to strain determines both Berry curvature and BCD. For As@MBP, the 4\% mechanical strain can increase the value of BCD to more than ten-times, form $2.1$ \AA \ to $22$ \AA, which is very significant for Borophosphene based 2D materials\cite{Joseph-awadesh}. 
On the other hand, As@AMBP retains a robust BCD and nonlinear Hall response across the investigated strain range, despite the strain-induced reduction in its band gap, making it a promising platform for strain-robust nonlinear Hall devices. This mechanical sensitivity indicates that the electronic properties of these monolayers can be tuned to enhance nonlinear transport without requiring strong spin-orbit coupling. Together, these results identify borophosphene derivatives as a promising platform for nonlinear electronic transport and piezoelectric-sensing applications.

 \acknowledgements
 RKT acknowledges the All India Council for Technical Education (AICTE) for providing the Doctoral Fellowship and financial support. This research was supported in part by the International Centre for Theoretical Sciences (ICTS) for the Frontiers in Quantum Condensed Matter Physics (code: ICTS/FQCMP2025/12).The authors also thank the High-Performance Computing facility at GITAM for providing the necessary computational resources for this work.

\bibliographystyle{apsrev4-2}
  \bibliography{main}
	
\end{document}


\begin{center}
\textbf{Supplemental Material for}\\[11pt]
\end{center}
    \title{\textcolor{blue}{Emergence of Berry curvature dipole and nonlinear Hall effect in a novel hexagonal borophosphene allotrope}
}
    
       \author{Ranjan Kumar Tung} 
      \thanks{}
      \affiliation{Department of Physics, GITAM School of Science, GITAM (Deemed to be University), Bengaluru 561203, Karnataka, India}
      \author{Dimple} 
      \thanks{dimple@gitam.edu}
      \affiliation{Theoretical and Computational Modeling of Materials Lab (TCM2), Department of Physics, GITAM School of Science, GITAM (Deemed to be University), Bengaluru 561203, Karnataka, India}
      \author{Venkatesan Subramanian}
      \thanks{}
      \affiliation{Department of Chemistry, Indian Institute of Technology (IIT) Madras,Chennai 600036, India}
      \affiliation{Centre for Advanced Computational and Theoretical Science (ACTS), SRM Institute of Science and Technology, Kattankulathur - 603203, Tamil Naidu, India }
       \affiliation{Department of Chemistry, SRM Institute of Science and Technology, Kattankulathur - 603203, Tamil Naidu, India }
      \author{Vijay Singh} 
      \thanks{vsingh2@gitam.edu}
      \affiliation{Department of Physics, GITAM School of Science, GITAM (Deemed to be University), Bengaluru 561203, Karnataka, India}

      \maketitle
The Supplemental Material provides comprehensive stability profiles for the optimized ground-state geometries 
of monolayer As@MBP, and As@AMBP are detailed in 
Figs.~S1--S3 and Tables~S1. The structural and electronic data for the uniaxial compressive strain study along the $x$ axis, applied from 0\% to 8\% in steps of 2\%. Tables~S2 and S3 list the bond lengths and bond angles of As@MBP and As@AMBP, respectively. In both systems, all bond lengths and the P--B--P and As--B--As angles decrease monotonically with strain. For As@MBP system, the B--B bond contracts from 1.60 to 1.55~\AA, and the P--B--P angle decreases from $125.05^\circ$ to $118.00^\circ$. In As@AMBP the As--B and P--B bonds shorten from 1.89 to 1.82~\AA\ and from 1.84 to 1.75~\AA, respectively. Fig.~S4 shows the low-energy Wannier-interpolated band structures of As@MBP together with the total Berry curvature $\Omega_z(\mathbf{k})$ (panels a--d) and the band-resolved Berry curvature contributions (panels e--h). The band gap systematically decreases from 66 to 38~meV as the strain increases from 2\% to 8\%. The Berry curvature remains sharply localized at the gap opening at the Dirac point between $\Gamma$--$X$, with opposite signs at the two nodes. Its peak magnitude reaches about $2\times10^{3}$~\AA$^{2}$ at 4\% strain and decreases to about $6\times10^{2}$~\AA$^{2}$ at 8\%. Fig.~S5 shows the BCD of As@AMBP as a function of energy relative to $E_F$ for the same strain values. The band gap decreases from 76 to 46~meV with increasing strain. The BCD displays alternating positive and negative peaks within $\pm0.15$~eV of $E_F$, with a magnitude of about 0.2~\AA\ that does not change significantly with strain, which confirms the robust response to applied strain discussed in the main text.

\begin{figure}
    \centering
    \includegraphics[width=\linewidth]{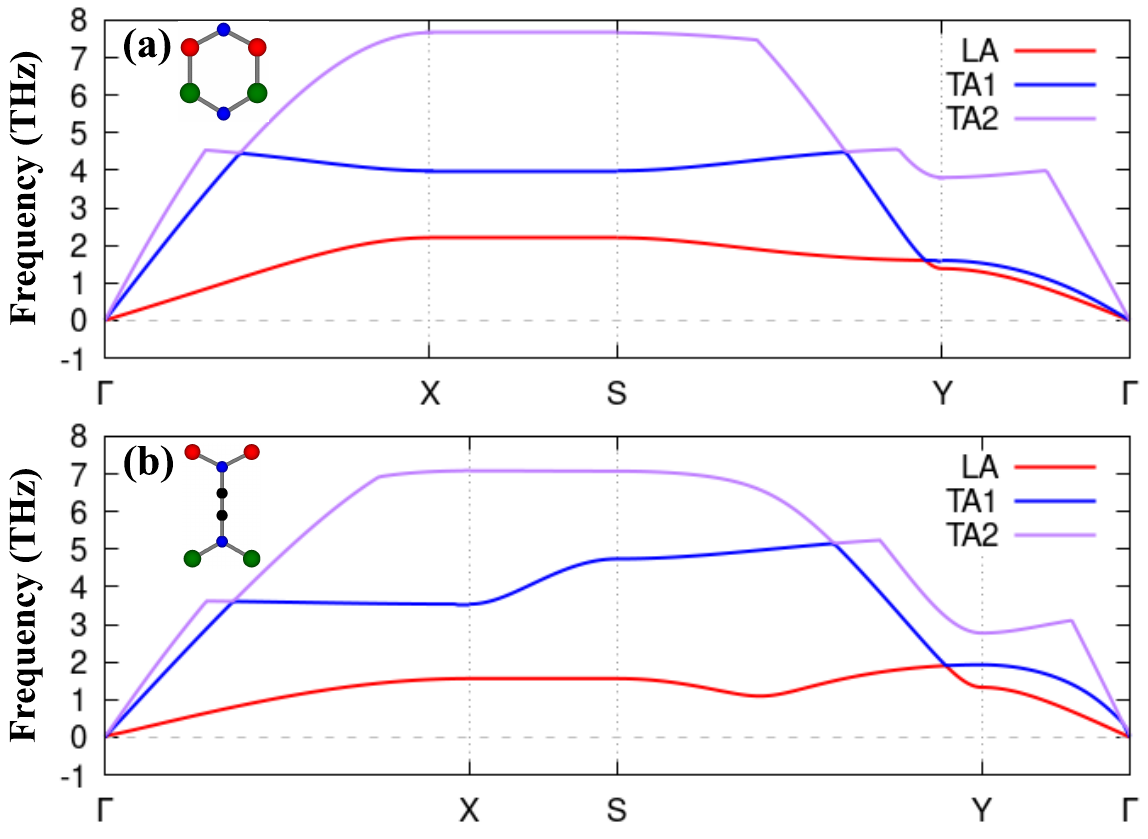}
    \caption{Acoustical phonon band structure for As@MBP (a) and As@AMBP (b); LA: longitudinal acoustic mode, TA1 and TA2 are the two transverse acoustic modes respectively}
    \label{fig:phonon}
\end{figure}

\begin{figure}
    \centering
    \includegraphics[width=\linewidth]{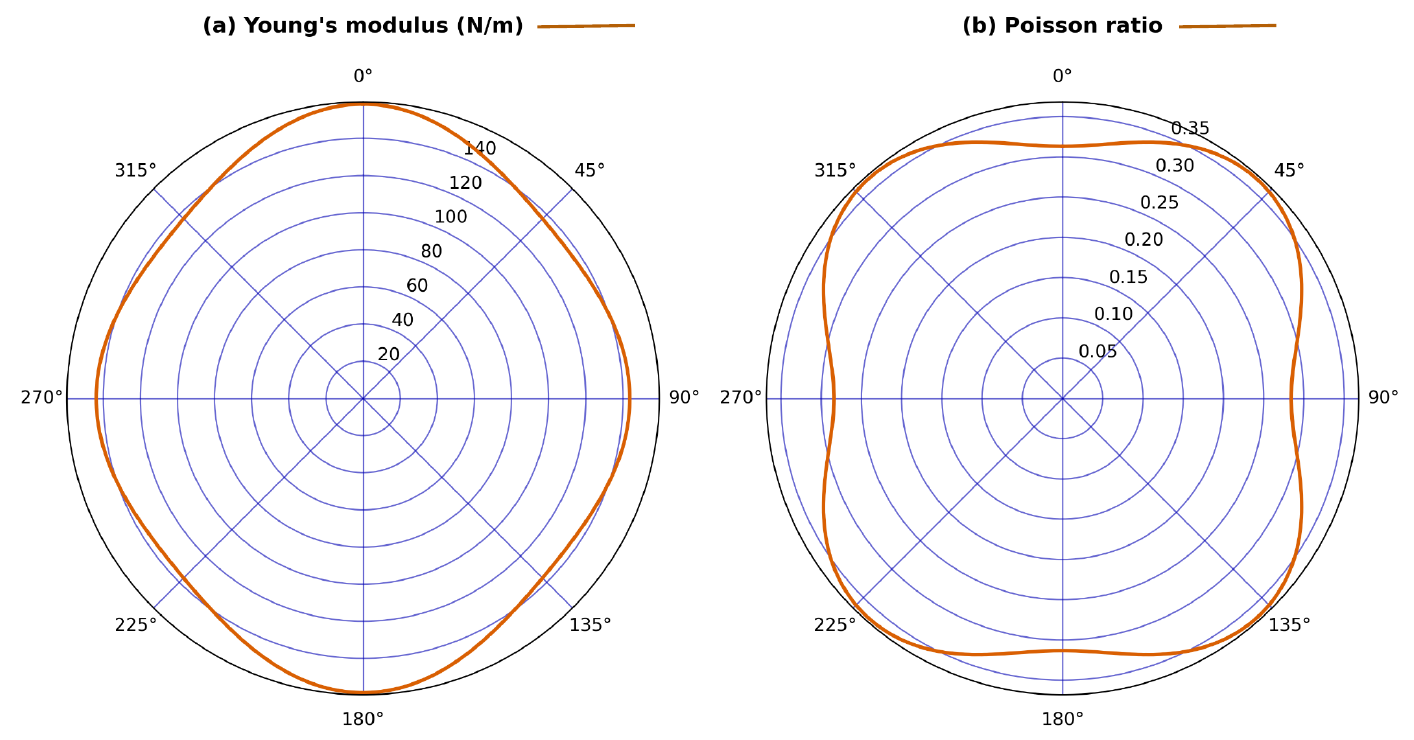}
    \caption{Polar diagrams of (a) the in-plane Young’s modulus and (b) the Poisson ratio for the As@MBP}
    \label{fig:placeholder}
\end{figure}

\begin{figure}
    \centering
    \includegraphics[width=\linewidth]{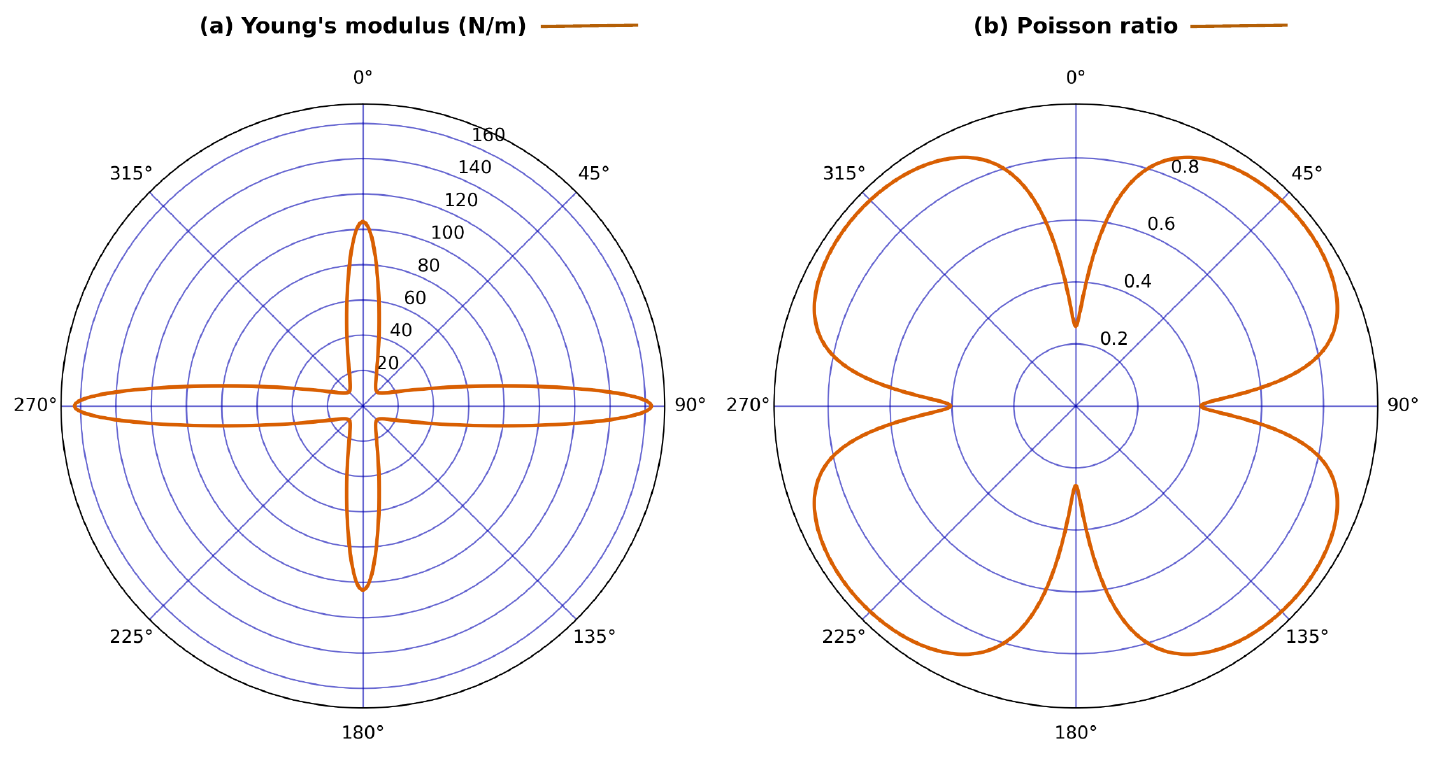}
    \caption{Polar diagrams of (a) the in-plane Young’s modulus and (b) the Poisson ratio for the As@AMBP}
    \label{fig:placeholder}
\end{figure}

\section{Elastic Stability and Anisotropic Mechanical Properties of 2D \text{As@AMBP} and \text{As@MBP}}

The mechanical stability and in-plane elastic behavior of monolayer $\text{As@AMBP}$ and $\text{As@MBP}$ were investigated based on their calculated stiffness tensors $[C]$ and compliance tensors $[S]$. For both 2D systems, the elastic constitutive relations are governed by four independent constants ($C_{11}$, $C_{22}$, $C_{12}$, and $C_{66}$):

\begin{equation}
[C]_{\text{As@AMBP}} = 
\begin{pmatrix}
116.40 & 47.08 & 0 \\
47.08 & 182.52 & 0 \\
0 & 0 & 2.91
\end{pmatrix} \text{N/m}, \quad
[S]_{\text{As@AMBP}} = 
\begin{pmatrix}
0.00959 & -0.00247 & 0 \\
-0.00247 & 0.00612 & 0 \\
0 & 0 & 0.34320
\end{pmatrix} \text{m/N}
\end{equation}

\begin{equation}
[C]_{\text{As@MBP}} = 
\begin{pmatrix}
173.93 & 49.42 & 0 \\
49.42 & 157.76 & 0 \\
0 & 0 & 50.16
\end{pmatrix} \text{N/m}, \quad
[S]_{\text{As@MBP}} = 
\begin{pmatrix}
0.00631 & -0.00198 & 0 \\
-0.00198 & 0.00696 & 0 \\
0 & 0 & 0.01994
\end{pmatrix} \text{m/N}
\end{equation}

\subsection{Mechanical Stability}
Both structures satisfy the 2D Born-Huang elastic stability criteria ($C_{11} > 0$, $C_{66} > 0$, and $C_{11}C_{22} - C_{12}^2 > 0$). Specifically, $C_{11}C_{22} - C_{12}^2$ evaluates to $19,028.5\text{ N}^2/\text{m}^2$ for $\text{As@AMBP}$ and $25,000.7\text{ N}^2/\text{m}^2$ for $\text{As@MBP}$. Furthermore, all stiffness matrix eigenvalues are strictly positive ($\lambda_1, \lambda_2, \lambda_3 = 2.91, 91.93, 206.98\text{ N/m}$ for $\text{As@AMBP}$ and $50.16, 115.77, 215.92\text{ N/m}$ for $\text{As@MBP}$), confirming intrinsic mechanical stability in both monolayers.

\subsection{Single-Crystal Anisotropy and Comparative Analysis}
The angle-dependent Young's modulus $Y(\theta)$ and Poisson's ratio $\nu(\theta)$ along an in-plane angle $\theta$ relative to the $x$-axis are expressed as:

\begin{equation}
Y(\theta) = \frac{1}{S_{11}\cos^4\theta + (2S_{12} + S_{66})\sin^2\theta\cos^2\theta + S_{22}\sin^4\theta}
\end{equation}

\begin{equation}
\nu(\theta) = -\frac{S_{12}\cos^4\theta + (S_{11} + S_{22} - S_{66})\sin^2\theta\cos^2\theta + S_{12}\sin^4\theta}{S_{11}\cos^4\theta + (2S_{12} + S_{66})\sin^2\theta\cos^2\theta + S_{22}\sin^4\theta}
\end{equation}

As summarized in Table~\ref{tab:mechanical_comparison}, the two materials exhibit contrasting elastic profiles:
\begin{itemize} 
    \item \textbf{As@MBP} exhibits near-isotropic elastic behavior with $Y(\theta)$ ranging tightly between $136.03\text{ N/m}$ and $158.45\text{ N/m}$ ($Y_{\max}/Y_{\min} = 1.165$). The near-unity Voigt-to-Reuss bounds ($K_V/K_R = 1.003$, $G_V/G_R = 1.007$) and small anisotropy index ($A_{\text{SU}} = 0.010$) confirm high structural homogeneity and orientation-independent elasticity.
    \item \textbf{As@AMBP} displays giant mechanical anisotropy driven by a soft shear constant ($C_{66} = 2.91\text{ N/m}$). This results in a minimal Young's modulus of $Y_{\min} = 11.30\text{ N/m}$ and a high Young's modulus anisotropy ratio ($Y_{\max}/Y_{\min} = 14.467$). Its Universal Elastic Anisotropy Index ($A_{\text{SU}} = 5.546$) and high $G_V/G_R$ ratio ($4.922$) further underline its strong directional dependence.
\end{itemize}

\begin{table}[h!]
\centering
\caption{Comparative single-crystal elastic bounds, polycrystalline averages (Voigt, Reuss, Hill), and anisotropy parameters for 2D $\text{As@AMBP}$ and $\text{As@MBP}$.}
\label{tab:mechanical_comparison}
\begin{tabular}{l c c c c c c}
\toprule
& \multicolumn{3}{c}{\textbf{As@AMBP}} & \multicolumn{3}{c}{\textbf{As@MBP}} \\
\cmidrule(lr){2-4} \cmidrule(lr){5-7}
\textbf{Single-Crystal Property} & \textbf{Min} & \textbf{Max} & \textbf{Anisotropy} & \textbf{Min} & \textbf{Max} & \textbf{Anisotropy} \\
\midrule
Young's Modulus $Y$ ($\text{N/m}$) & 11.300 & 163.479 & 14.467 & 136.034 & 158.449 & 1.165 \\
Shear Modulus $G$ ($\text{N/m}$)   & 2.914  & 48.411  & 16.614 & 50.164  & 58.062  & 1.157 \\
Poisson's Ratio $\nu$              & 0.258  & 0.941   & 3.648  & 0.284   & 0.364   & 1.282 \\
\midrule
\textbf{Polycrystalline Moduli} & \textbf{Voigt} & \textbf{Reuss} & \textbf{Hill} & \textbf{Voigt} & \textbf{Reuss} & \textbf{Hill} \\
\midrule
Bulk Modulus $K$ ($\text{N/m}$)  & 98.267 & 92.930 & 95.598 & 107.629 & 107.348 & 107.489 \\
Shear Modulus $G$ ($\text{N/m}$) & 27.053 & 5.497  & 16.275 & 54.189  & 53.825  & 54.007  \\
Bulk/Shear Ratio $K/G$           & 3.632  & 16.906 & 5.874  & 1.986   & 1.994   & 1.990   \\
\midrule
\textbf{Anisotropy Metrics} & \multicolumn{3}{c}{$A_{\text{SU}} = 5.546, \, G_V/G_R = 4.922$} & \multicolumn{3}{c}{$A_{\text{SU}} = 0.010, \, G_V/G_R = 1.007$} \\
\bottomrule
\end{tabular}
\end{table}
    \begin{table}[h!]
\centering
\caption{Bond lengths and bond angles of As@MBP under uniaxial compressive strain along the $x$ axis.}
\label{tab:As_MBP}
\begin{tabular}{cccccc}
\hline
 & \multicolumn{5}{c}{\textbf{Amount of uniaxial strain along x-axis}} \\ \cline{2-6}
\textbf{Structural Parameter} & \textbf{0\%} & \textbf{2\%} & \textbf{4\%} & \textbf{6\%} & \textbf{8\%} \\ \hline
$P - B_2$ (\AA) & 1.81& 1.79& 1.77& 1.75& 1.73\\
$As - B_1$ (\AA) & 1.88& 1.86& 1.84& 1.82& 1.80\\
$As - As$/ $P - P$ (\AA)& 3.22& 3.16& 3.09& 3.03& 2.96\\
$B - B$ (\AA)& 1.60& 1.59& 1.58& 1.56& 1.55\\
$\angle P - B - P$ ($^\circ$) & 125.05& 123.60 & 121.81 & 120.00 & 118.00 \\
$\angle As - B - As$ ($^\circ$) & 117.89& 116.19 & 114.19 & 112.29 & 110.28 \\ \hline
\end{tabular}
\end{table}


\begin{figure}
    \centering
    \includegraphics[width=\linewidth]{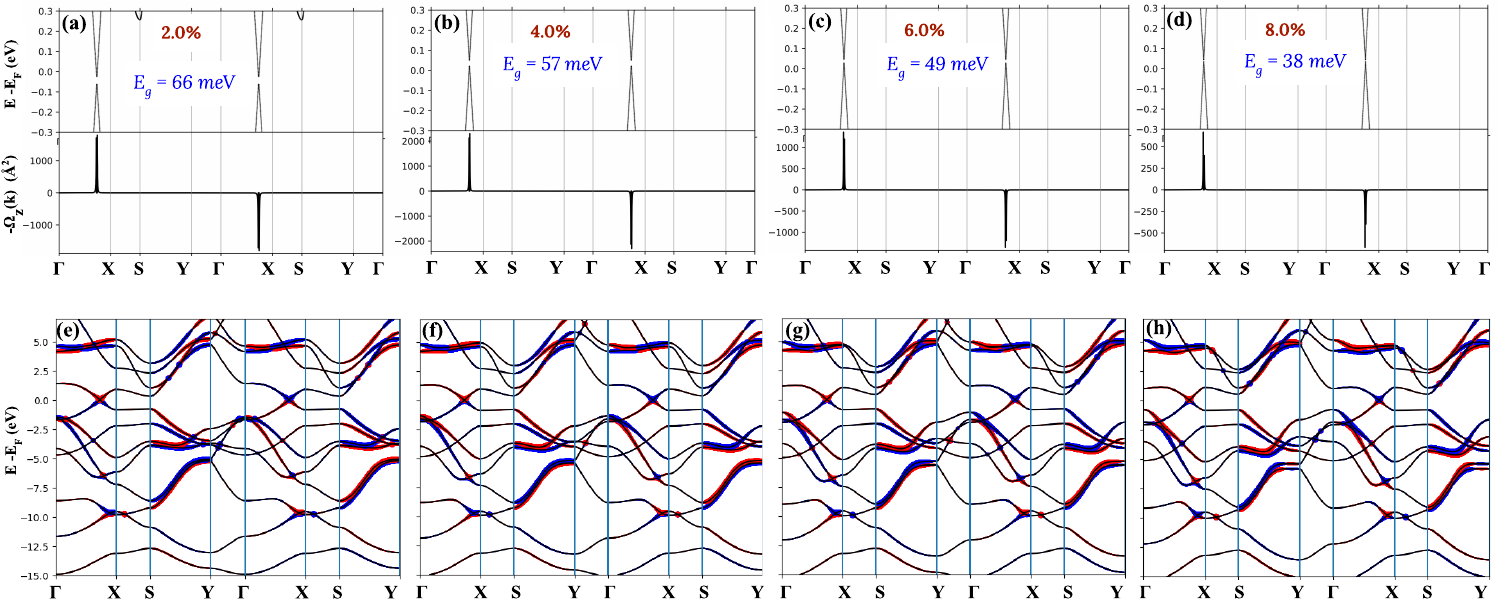}
    \caption{Effect of uniaxial compressive strain along the $x$ axis on As@MBP for strains of 2\%, 4\%, 6\%, and 8\% (left to right). (a)--(d) Low-energy Wannier-interpolated band structures with the total Berry curvature $-\Omega_z(\mathbf{k})$ (in \AA$^2$) plotted below, together with the band gap $E_g$ at each strain. (e)--(h) Corresponding band-resolved Berry curvature contributions in the first Brillouin zone, with red and blue denoting positive and negative values, respectively.}
    \label{fig:placeholder}
\end{figure}

\begin{table}[h!]
\centering
\caption{Bond lengths and bond angles of As@AMBP under uniaxial compressive strain along the $x$ axis.}
\label{tab:As_AMBP}
\begin{tabular}{cccccc}
\hline
 & \multicolumn{5}{c}{\textbf{Amount of uniaxial strain along x-axis}} \\ \cline{2-6}
\textbf{Structural Parameter} & \textbf{0\%} & \textbf{2\%} & \textbf{4\%} & \textbf{6\%} & \textbf{8\%} \\ \hline
$As - B$ (\AA) & 1.89& 1.87& 1.85& 1.83& 1.82\\
$P - B$ (\AA) & 1.84& 1.81& 1.79& 1.77& 1.75\\
$As - As$ (\AA) & 3.28& 3.21& 3.15& 3.08& 3.02\\
$P - P$ (\AA) & 3.28& 3.21& 3.15& 3.08& 3.02\\
$\angle As - B - As$ ($^\circ$) & 119.98& 118.15 & 116.20 & 114.18 & 112.10 \\
$\angle P - B - P$ ($^\circ$) & 126.43& 124.86 & 123.13 & 121.28 & 119.34 \\ \hline
\end{tabular}
\end{table}  

\begin{figure}
    \centering
    \includegraphics[width=0.8\linewidth]{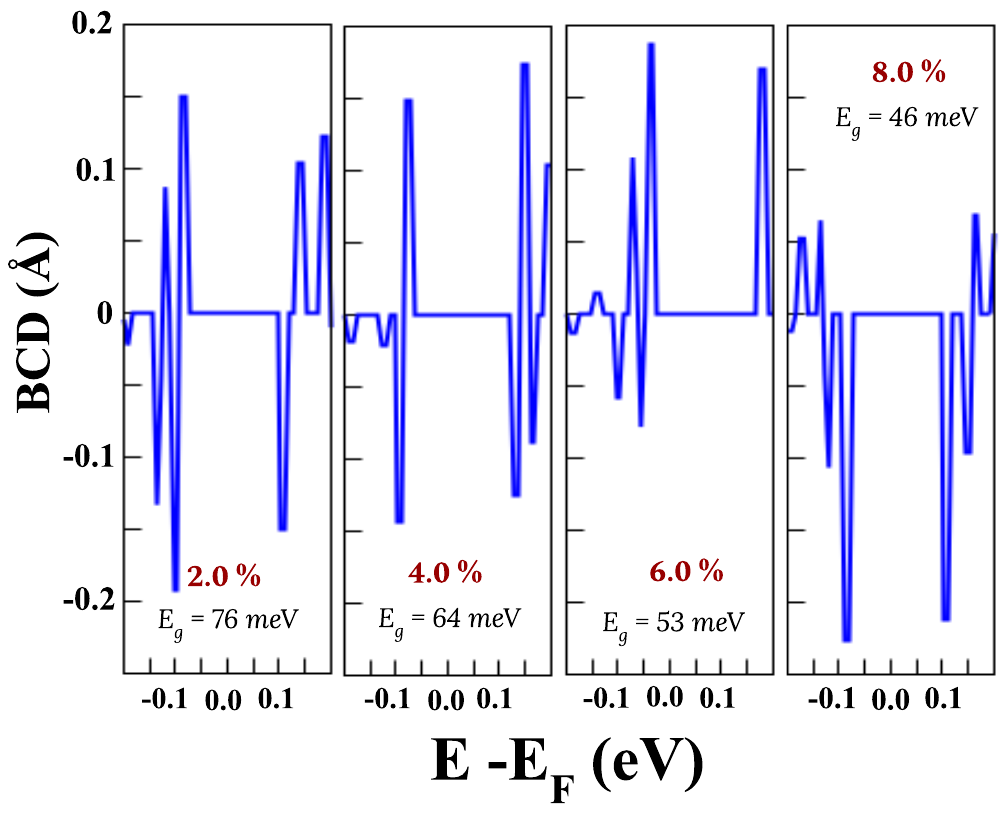}
    \caption{Berry curvature dipole (BCD) of As@AMBP as a function of energy relative to the Fermi level $E_F$ under uniaxial compressive strain along the $x$ axis of 2\%, 4\%, 6\%, and 8\% (left to right). The band gap $E_g$ decreases from 76 to 46~meV with increasing strain, while the magnitude of the BCD remains comparable.}
    \label{fig:placeholder}
\end{figure}